\documentclass[fleqn,usenatbib,onecolumn]{mnras}

\usepackage{newtxtext,newtxmath}

\usepackage[T1]{fontenc}
\usepackage{ae,aecompl}
\usepackage{amsmath}

\usepackage{graphicx}	
\usepackage{amsmath}	
\usepackage{amssymb}	
\usepackage{multicol}   
\usepackage{bm}		    
\usepackage{pdflscape}	
\usepackage{array}
\usepackage{hyperref}
\usepackage{xspace}                         

\usepackage[T1]{fontenc}
\usepackage{ae,aecompl}

\newcommand{\ox}{\ensuremath{\mathrm{O}}}
\newcommand{\oz}{\ensuremath{\mathrm{O}_3}}
\newcommand{\hy}{\ensuremath{\mathrm{H}}}

\newcommand{\oxod}{\ensuremath{\ox(^{1}\mathrm{D})}}

\newcounter{chem}
\newcounter{temp}
\newenvironment{chequation}{%
  \setcounter{temp}{\value{equation}}%
  \setcounter{equation}{\value{chem}}%
}{%
  \setcounter{chem}{\value{equation}}%
  \setcounter{equation}{\value{temp}}%
}

\newcommand{\ang}[1]{\ensuremath{{#1}^{\circ}}}

\defcitealias{Boutle2017}{Paper~I}

\title[Ozone chemistry on tidally locked M dwarf planets]{Ozone chemistry on tidally locked M dwarf planets}

\author[J. S. Yates et al.]{Jack S. Yates,$^{1,2}$,
Paul I. Palmer$^{1,2}$\thanks{E-mail: pip@ed.ac.uk},  
James Manners$^{3}$,
Ian Boutle$^{3}$,
Krisztian Kohary$^{4}$,
\newauthor
Nathan Mayne$^{4}$,
Luke Abraham$^{5,6}$
\\
$^{1}$ School of GeoSciences, University of Edinburgh, King's Buildings, Edinburgh, EH9 3FF, UK \\
$^{2}$ Centre for Exoplanet Science, University of Edinburgh, Edinburgh, UK\\
$^{3}$ Met Office, Exeter, EX1 3PB, UK \\
$^{4}$ Astrophysics Group, University of Exeter, Exeter, EX4 2QL, UK \\
$^{5}$ National Centre for Atmospheric Science, University of Cambridge, CB2 1EW, UK \\ 
$^{6}$ Department of Chemistry, University of Cambridge, Cambridge, CB2 1EW, UK\\
}

\date{Accepted XXX. Received YYY; in original form ZZZ}

\pubyear{2019}

\begin{document}
\label{firstpage}
\pagerange{\pageref{firstpage}--\pageref{lastpage}}
\maketitle

\begin{abstract}
We use the Met Office Unified Model to explore the potential of a
tidally locked M dwarf planet, nominally Proxima Centauri b irradiated
by a quiescent version of its host star, to sustain an atmospheric
ozone layer. We assume a slab ocean surface layer, and an Earth-like
atmosphere of nitrogen and oxygen with trace amounts of ozone and
water vapour. We describe ozone chemistry using the Chapman mechanism
and the hydrogen oxide (HO$_x$, describing the sum of OH and HO$_2$)
catalytic cycle. We find that Proxima Centauri radiates with
sufficient UV energy to initialize the Chapman mechanism. The result
is a thin but stable ozone layer that peaks at 0.75 parts per million
at 25~km. The quasi-stationary distribution of atmospheric ozone is determined by
photolysis driven by incoming stellar radiation and by atmospheric
transport. Ozone mole fractions are smallest in the lowest 15~km of
the atmosphere at the sub-stellar point and largest in the nightside
gyres. Above 15~km the ozone distribution is dominated by an
equatorial jet stream that circumnavigates the planet. The
nightside ozone distribution is dominated by two cyclonic Rossby
gyres that result in localized ozone hotspots.  On the dayside the atmospheric lifetime is
determined by the HO$_x$ catalytic cycle and deposition to the
surface, with nightside lifetimes due to chemistry much longer than
timescales associated with atmospheric transport. Surface UV values
peak at the substellar point with values of 0.01~W/m$^2$, shielded by
the overlying atmospheric ozone layer but more importantly by water
vapour clouds. 
\end{abstract}





\section{Introduction}

We are only just beginning to classify the rapidly growing number of
extrasolar planets (exoplanets) that orbit stars outside our solar
system. In the absence of other reference points, this classification
exercise is guided by the mass and radii of solar system planets, even
though they may not share similar evolutionary pathways
\cite[]{Seager2007,Spiegel2014,Weiss2014}. Exoplanets range from
scientific curiosities \cite[e.g. COROT-Exo-7b,][]{Fressin2009} to
candidates for supporting life
\citep{Quintana2014,Anglada-Escude2016,Gillon2017}. Using ever more
sophisticated technology, we are now detecting Earth-like exoplanets
that receive similar amounts of stellar irradiation as Earth receives
from the Sun, placing them into  the circumstellar habitable zone
(CHZ).  From an anthropocentric perspective, the CHZ is a region
around the host star that would support liquid water on the planetary
surface, a requirement for life \citep{Kasting1988, Kasting1993};
meeting this criterion means a planet is habitable and not that it is
necessarily inhabited. A zeroth order estimate of the CHZ can be
determined from the stellar irradiance and the star-planet
distance. However, planetary atmospheres play a significant role in
the planetary energy balance depending on atmospheric composition,
e.g. infrared absorbers such as greenhouse gases \cite[]{Meadows2018a}
and reflecting and absorbing aerosol particles and clouds.
Consideration of a planetary atmosphere is therefore integral for
understanding whether an exoplanet is habitable. Here, we use the
computational framework from a leading 3-D Earth system model to
describe the interplay between atmospheric dynamics and a simplified
description of atmospheric ozone chemistry on a tidally-locked M dwarf
planet, nominally Proxima Centauri b \citep{Anglada-Escude2016}.  The
planet is irradiated by a M dwarf spectrum that is representative of
an older, quiescent version of its host star. We use 3-D model
calculations to study variations in atmospheric ozone and to
understand how ozone affects the habitability of the planet, e.g.
surface UV environment and broader impacts on climate.
  
The first detections of planets around main-sequence stars came in
1995 \citep{Mayor1995}. In the 2000s, automated surveys revolutionised
the field yielding vast quantities of planet detections
(e.g. \cite{Borucki2016}). To date, thousands of exoplanets have been
discovered \citep[e.g.][]{Morton2016, Thompson2018} across a broad
range of bulk planetary properties. Telescope and detector
technologies have already allowed characterisation of giant exoplanet
atmospheres through transmission (e.g., \cite{Sing2011}) and emission
(e.g., \cite{Todorov2014}) spectroscopy and direct imaging
\cite[]{Lagrange2010}. The first detection of an exoplanet atmosphere
was achieved using transmission spectroscopy and showed
absorption by the sodium doublet in the atmosphere of the hot Jupiter
HD~209458b \cite[]{Charbonneau2002}. 
Recent surveys
have detected Earth-sized planets on close orbits around cold M dwarf
stars, notably Proxima Centauri b \citep{Anglada-Escude2016} and the
TRAPPIST 1 planets \citep{Gillon2017}. For these planets, orbital
distance and low stellar flux compensate for each other, resulting in
Earth-like levels of top-of-the-atmosphere radiation, placing them in
the CHZ \citep{Kasting1993}.

It is no coincidence that these planets have thus far been detected
only around M dwarf stars. These stars represent 70\% of all stars
\citep{Bochanski2010}, and the detectability of their orbiting planets
is greater due to the increased planet-to-star radius ratio. The
detectability of a planet also increases as the orbital semi-major
axis decreases because of the diminished effect of inclination and the
subsequent increased frequency of transits and the larger
radial-velocity signature. Demographic studies suggest that M dwarfs
are more likely than larger stars to host small planets
\citep{Shields2016}. Collectively, these observational factors suggest
that M dwarfs are our best hope for finding exoplanets in the
CHZ. However, there are many unknowns in the habitability assessment
of a candidate M dwarf, e.g. the manifold impacts of climate on
stellar and planetary environments \citep{Shields2016}. Until
telescopes are capable of detailed observations of M dwarf planets, we
are limited to examining them through investigative modelling efforts
and focus on the influence of atmospheric chemical composition.

Previous studies using general circulation models (GCMs) have shown
that close-in M dwarf planets, tidally locked or otherwise, may be habitable
\citep[e.g.][]{Boutle2017,Kopparapu2017,Turbet2017,Meadows2018}. For a
terrestrial planet the global circulation is driven almost entirely by
stellar radiation, which is by far the largest heat source in the
system. Naturally this creates spatial inhomogeneities. For a
tidally-locked planet with a permanent day and night side, these
inhomogeneities are even more exaggerated. In this scenario, the
3-dimensional nature of the system -- a hot day side, a cold night
side and an atmospheric jet transporting heat and moisture around the equator --
means that 1-dimensional models are not necessarily sufficient to
properly simulate the atmospheric dynamics. This is especially true
when clouds and convection are involved, as these processes are
difficult to simulate in 1 dimension \citep[see e.g.][for a discussion
  of these difficulties]{Tompkins2000,Palmer2012}. 
Historically, GCM studies of M dwarf planets have not incorporated an
interactive chemistry scheme, partly due to the uncertainties in the
atmospheric composition, partly due to the perceived sufficiency of
understanding atmospheric chemistry using 1D models, and partly
because of the perception that it plays a insignificant contribution
to the mean climate. Informed by the results from GCMs, one might
expect that the spatial inhomogeneities in the circulation and
dynamics drives non-uniform chemistry, resulting in planets that have
different day- and nightside chemical environments
\citep{Proedrou2016,Drummond2018}. This is relevant to tidally locked
planets, which have: broad Rossby gyres on the night side
\citep{Showman2011} that temporarily trap air that is consequently
subject to extended periods of radiative cooling, resulting in
localized cold spots; extreme day-night temperature differences
compared with a fast-rotating planet like Earth; and portions of the
atmosphere or surface that are never exposed to stellar radiation.

Integrating atmospheric chemistry and physics into GCMs is therefore a
natural next step for exoplanet science, and is relevant also for
understanding and testing potential atmospheric biosignatures
\citep{Seager2014}. Spatial inhomogeneities in atmospheric
biosignatures, driven by atmospheric chemistry and physics, will
affect our ability to detect them (e.g. transmission geometry) and to
interpret them in the absence of 3-D models.
In the case of habitable M-dwarf planets, direct detection, where
spectra are hemispherically averaged, is not likely to be possible for
many years due to their associated small angular  separations from
their host star. Transmission and thermal phase curves therefore
represent the main current observing modes for these planets, for
which terminator profiles and day-night contrasts will be the most
important.
Conversely, the observed
variations in these biosignatures, interpreted using 3-D model
analogues, may identify which ones are most robust.

Previous work that studied a tidally locked Earth, although orbiting
the Sun in 365 days, showed significant chemical differences on the
day and night side of the planet \citep{Proedrou2016}. Using a simple
ozone photochemistry model, they showed that ozone was present on both
sides but was present in higher concentrations on the night side. They
also found that for the night side, where chemical e-folding lifetimes
are longest, atmospheric transport was most important for determining
ozone distributions. This work only has limited applicability to M
dwarf planets such as Proxima Centauri b. The stellar radiation
spectrum plays a critical role in determining ozone chemistry. It is
unclear {\em a~priori} whether a star emitting less UV radiation than
the Sun would be able to establish and sustain an ozone layer.
Generally, being able to maintain a layer of radiatively active gas,
such as ozone, will influence the habitability of the planet directly
via the surface radiation environment and indirectly via climate
feedbacks. Recent work has used the CAM-Chem 3-D model to study of
atmospheric chemistry on M dwarfs \citep{Chen2018}. These authors also
discussed the day-night differences in chemistry, and emphasized the
importance of using self-consistent chemistry fields for interpreting
observations. We discuss this study in the context our work later in
section \ref{sect:discuss}.

In this work we examine the sustainability of an ozone layer on a
tidally locked M dwarf planet. We use a global 3-D GCM that describes
a terrestrial exoplanet, nominally Proxima Centauri b, building on
\cite{Boutle2017}. The GCM is a generalized version of the Met Office
Unified Model (UM), which is used extensively for short-term weather
prediction and long-term climate studies for the Earth
\cite[]{Walters2017}. The UM has been used previously to study the
atmospheric physics of terrestrial exoplanets and hot Jupiters
(e.g. \cite{Mayne2014,Mayne2014b,Boutle2017,Lewis2018,Drummond2018}). We
couple this model with the Chapman mechanism \cite[]{Chapman1930} that
describes ozone chemistry in Earth's stratosphere with the hydrogen
oxide catalytic cycle. The UM and the atmospheric chemistry are
described in Section 2. In Section 3, we report our results that
describe variations of atmospheric ozone on a tidally-locked M dwarf
planet and the responsible chemical reactions. We conclude the paper
in Section 4.

\section{Model Description}
\label{sect:model}

Table \ref{tab:planetparams} shows an overview of the physical
parameters that define the Proxima Centauri b simulation, following
\cite{Boutle2017}. For the sake of brevity, we focus here on key
aspects of the simulation and refer the reader to
\cite{Mayne2014,Mayne2014b,Boutle2017} for a comprehensive model
description. We run the model at a horizontal spatial resolution of
$2^{\circ}$ (latitude) $\times 2.5^{\circ}$ (longitude), with the
substellar point defined at latitude and longitude \ang{0}. We
increase the upper altitude boundary of the model from 40~km
\cite[]{Boutle2017} to 85~km, described on 60 levels of which 38
levels describe the atmosphere from the surface to 40 km and are
identical to those used by \cite{Boutle2017}. We increased the
model upper boundary to improve the description of ozone dynamics in the
upper atmosphere.  We also reduced the dynamical model time step from
20 minutes, as adopted by \cite[]{Boutle2017}, to 12 minutes to avoid high
wind speeds above altitudes of 40~km violating the Courant-Friedrichs-Lewy
condition. Chemistry and radiation timesteps were left unchanged from
a default value of one hour. 

\begin{table}
    \centering
    \begin{tabular}{lc}
      Parameter & Value \\
      \hline
      Semi-major axis / AU            & 0.0485  \\
      Stellar irradiance / W m$^{-2}$ & 881.7  (0.646)\\
      Orbital period / Earth days    & 11.186 \\
      $\Omega$ /rad s$^{-1}$          & 6.501 $\times 10^{-6}$ \\
      Eccentricity                   & 0    \\
      Obliquity                      & 0     \\
      Radius / km                    & 7160 (1.1)        \\
      $g$ / m s$^{-1}$                & 10.9 \\ 
    \end{tabular}
    \caption{Planetary parameters for the simulated Proxima Centauri b planet. Numbers in round brackets
      represent the ratio with resepect to values for Earth.} 
    \label{tab:planetparams}
\end{table}

\begin{figure}[h!]
    \centering
    \includegraphics[width=0.6\columnwidth]{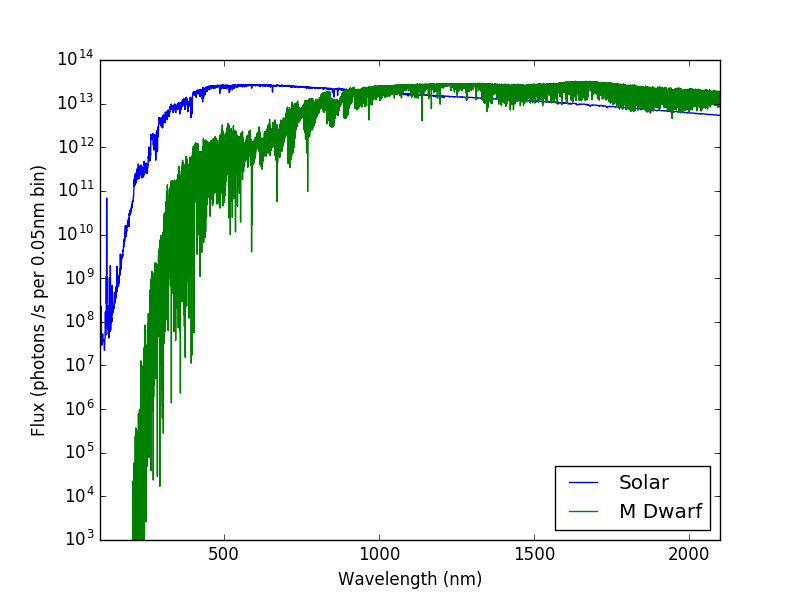}
    \caption{Top-of-the-atmosphere fluxes (photons/s/0.05~nm bin) as a function
      of wavelength (nm) for Proxima Centauri b (green)
      and Earth (blue), both scaled to the semi-major axis of orbit.}
    \label{ch3:fig:spectrum}
\end{figure}

Figure \ref{ch3:fig:spectrum} shows our assumed radiation spectrum for
Proxima Centauri, taken from BT-Settl model library for an M dwarf
with $T_\mathrm{eff} = 3000~\mathrm{K}$, $g = 1000~\mathrm{ms^{-2}}$
and $\mathrm{metallicity} = 0.3~\mathrm{dex}$ \citep{Rajpurohit2013}.
Using this spectrum, which approximately follows a blackbody
distribution, allows us to directly compare our results with previous
results that used fixed atmospheric composition \cite[]{Boutle2017}.
Consequently, Proxima Centauri emits less radiation at UV wavelengths
relevant to the Chapman mechanism (see photochemical reaction
\ref{eq:chapman} below) than the Sun because of its lower blackbody
temperature. We acknowledge that this energy distribution does not
include stellar activity, so represents a quiescent star, and does not
include Lyman-$\alpha$ emission.  Numerical experiments by the authors
(not shown) suggest that including Lyman-$\alpha$ emissions would
likely result in only a small increase in total O$_3$ columns and a
small corresponding change in atmospheric heating.

The UM has the capability to describe gas
\cite[]{Morgenstern2013,Telford2013} and aerosol phase
\cite[]{Mann2010} chemical mechanisms, developed collectively under
the UK Chemical and Aerosol (UKCA) project,  that is typically used to
explore how atmospheric composition affects Earth's climate. Here, we
use relatively simple ozone chemistry (described below), but the
underlying infrastructure allows this work to be easily extended to
study more complex organic chemistry.  The atmospheric chemistry is
fully 3-D with advection and turbulent mixing, and chemical species are
included in the radiation scheme, making the model
self-consistent.

We use the Chapman mechanism (reactions R1--R4, \cite{Chapman1930}) to
describe the main characteristics of atmospheric ozone on our
tidally-locked M dwarf. This mechanism forms the basis of our
understanding of observed variations in stratospheric ozone. However,
the mechanism overpredicts stratospheric ozone on Earth mainly because
it does not take into account catalytic cycles. The mechanism is
initiated in R1 by the photolysis of molecular oxygen by high-energy
UV photons ($\lambda <$240~nm), producing atomic oxygen in the
ground-level triplet state O$(^3P)$, denoted here as O, which are
highly reactive and combine rapidly with O$_2$ to form ozone (R2). $H$
denotes an inert molecule such as N$_2$ or O$_2$ that dissipates the
energy associated with the reaction. The resulting ozone molecules can
be photolyzed (R3) at longer wavelengths (less energetic photons,
$\lambda <$320~nm) to produce O in its excited singlet state O$(^1D)$
that is rapidly stabilized to O$(^3P)$ by collision with N$_2$ or
O$_2$. R4 describe the slow loss of ozone due to reaction with atomic
oxygen. 
\begin{chequation}
  \begin{eqnarray}
\ox_{2} + h\nu    \rightarrow  & \ox + \ox        &\left(\lambda < 240 \mathrm{nm}\right)  \\
\ox + \ox_2 + H   \rightarrow & \oz + H         & \\
\oz + h \nu       \rightarrow & \ox_2 + \ox  &\left( \lambda < 320 \mathrm{nm}\right)  \\
\oz + \ox        \rightarrow & \ox_2 + \ox_2   &
\end{eqnarray}
\label{eq:chapman}
\end{chequation}

We include the HO$_x$ (sum of the hydroxyl radical OH and hydroperoxy radical HO$_2$) catalytic cycle
\cite[]{Roney1965}, acknowledging that our atmosphere also includes
water vapour that can lead to ozone loss.
\begin{chequation}
\begin{eqnarray}
\hy_{2}\ox + \oxod \rightarrow && 2 \ox\hy          \\ \ox \hy + \oz
\rightarrow && \hy \ox_2 + \ox_2 \\  \hy \ox_2 + \oz     \rightarrow &&
\ox \hy + 2 \ox_2 \\ {\bf Termination: \hspace{1cm}} \ox \hy + \hy
\ox_2   \rightarrow && \hy_{2}\ox + \ox_2 
\end{eqnarray}
\label{eq:catalytic}
\end{chequation}
The catalytic cycle comprises of an initiation step (R5), propagation
steps that destroy ozone and recycle the catalysis (R6 and R7), and a
termination step that represents a loss of the catalysis (R8). We also
retain values used by UKCA for ozone dry deposition, which represents
a minor loss terms except near the planetary surface.  We describe dry
deposition of ozone over water following \cite{Giannakopoulos1999},
using a value of 0.05~cm~s$^{-1}$ \cite[]{Ganzeveld1995}.

We define our initial conditions using an Earth-like configuration,
comprised of $\mathrm{N}_2$ (approximately 77\%), $23.14\%$
$\mathrm{O}_2$ (mass mixing ratio, MMR) and $0.0594 \%$
$\mathrm{C}\mathrm{O}_2$ (MMR), with the remainder being composed of
ozone chemistry tracers that we describe above. We do not consider CH$_4$
and N$_2$O in our atmosphere because these trace gases imply
life and significantly complicate our chemical mechanism, but they are
included in \cite{Boutle2017}. With the
exception of N$_2$, CO$_2$, $\ox_2$ and $\hy_{2}\ox$, we set the other gases
associated with chemistry to a uniformly distributed MMR value of
$10^{-9}$ to minimize the influence of initial conditions on
subsequent model fields; $\hy_{2}\ox$ is calculated based on evaporation 
from the slab ocean. We spin-up the model fields for 20 years from
those initial conditions. We find that the atmospheric ozone state reaches
a quasi equilibrium after 5--10~years subject to stochastic changes in
atmospheric dynamics. As a conservative approach, the results we present
here are a mean of 120~days that immediately follow the 20-year spin-up
period. We perform three model runs: a calculation including the Chapman
O$_3$ mechanism with and without the HO$_x$ catalytic cycle, and a model
run using a fixed Earth-like O$_3$ distribution, following \cite{Boutle2017}.

We use two separate but consistent radiation schemes in the model. The
SOCRATES radiative transfer model describes the radiation processes in
the physical model using six short-wave and nine long-wave bands
\cite[]{Edwards1996,Manners2015}, and represents the default UM
radiation scheme. The chemistry module uses a separate radiation
scheme, Fast-JX \cite[]{Wild2000,Telford2013}, that is spectrally
resolved into 18 bands (Table \ref{tab:bands}) corresponding to
atmospheric chemistry. 

\begin{table}
    \centering
    \begin{tabular}{cc}
Wavelength band (nm) & TOA flux at 1 AU (photons s$^{-1}$ cm$^{-2}$)
\\\hline 187     & $2.213 $     \\ 191     & $2.874 $     \\ 193     &
$3.589 $     \\ 196     & $1.205 \times 10^{1}$     \\ 202     &
$3.502 \times 10^{1}$     \\ 208     & $5.388 \times 10^{2}$
\\ 211     & $6.885 \times 10^{2}$     \\ 214     & $1.684 \times
10^{5}$     \\ 261     & $2.107 \times 10^{7}$     \\ 267     & $4.660
\times 10^{7}$     \\ 277     & $1.922 \times 10^{8}$     \\ 295     &
$9.886 \times 10^{8}$     \\ 303     & $1.066 \times 10^{9}$
\\ 310     & $3.349 \times 10^{9}$     \\ 316     & $1.390 \times
10^{10}$     \\ 333     & $1.110 \times 10^{11}$     \\ 380     &
$1.980 \times 10^{12}$     \\ 574     & $1.319 \times 10^{15}$
    \end{tabular}
    \caption{M dwarf TOA fluxes at spectral bands used by the UKCA model, reported at 1 AU.}
    \label{tab:bands}
\end{table}


\section{Results}

Here, we report results from our numerical experiments that follow our
20-year spin-up period, as described above. Reported values represent
a 120-day mean. We focus our attention on ozone chemistry from our
control run, referring the reader to \cite{Boutle2017} for a detailed
discussion on the corresponding atmospheric physics and
climate. Sensitivity runs will be discussed explicitly with reference
to the control run.

\subsection{Hemispheric Mean 1-D Meteorological and Ozone Structure}

\begin{figure}
    \centering
    \includegraphics[width=\textwidth]{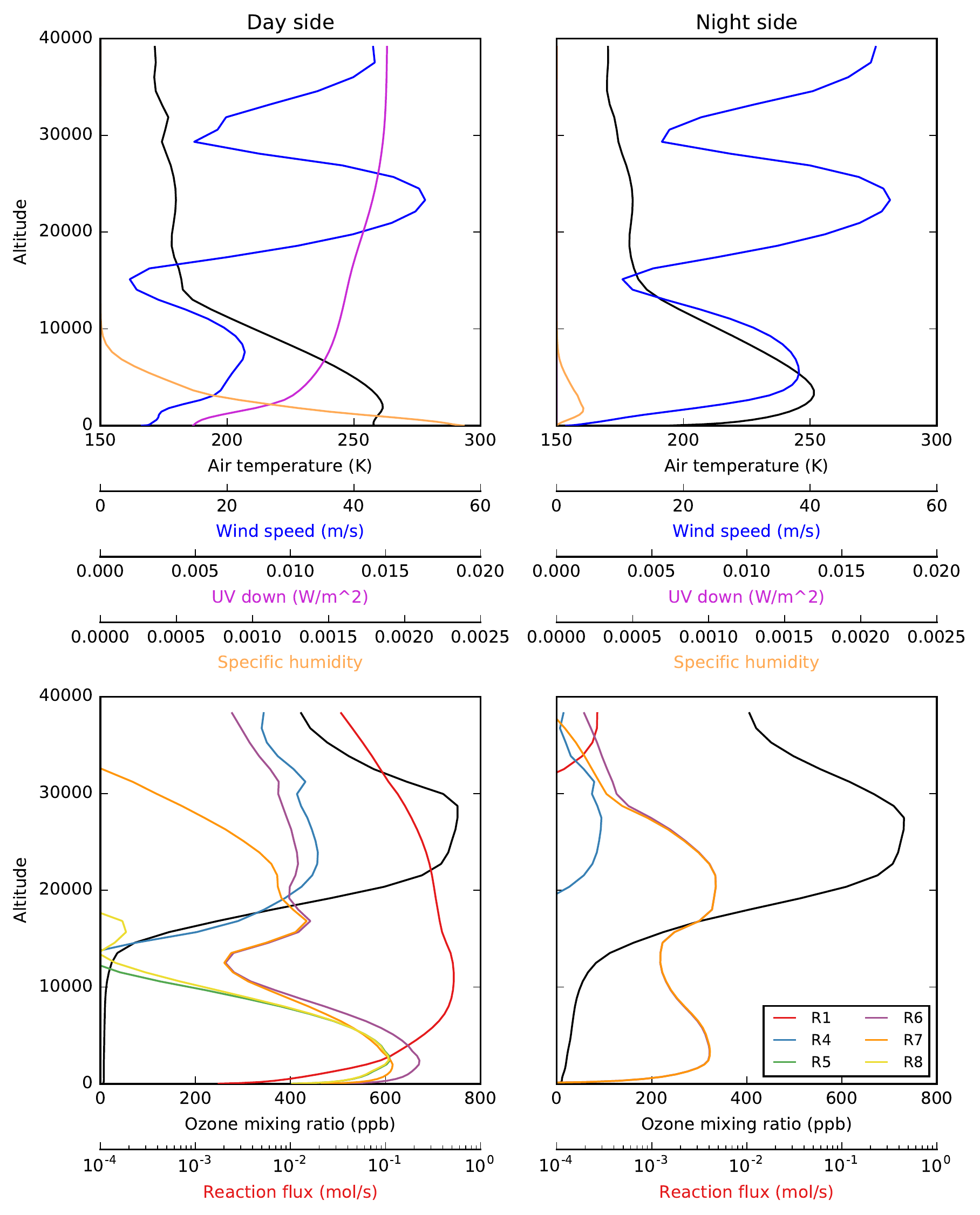}
    \caption{(Left panels) Dayside and (Right panel) nightside
      hemispheric means of (Top panels) meteorological and (Bottom
      panels) chemical parameters on a tidally-locked M dwarf
      planet. Values correspond to a 120-day mean immediately
      succeeding a 20-year spin-up period.  Black lines in the bottom
      panels show ozone concentration (ppb) per model grid box and the
      coloured lines show the corresponding reaction fluxes (mol/s),
      which are plotted on a logarithmic scale as flux per model grid
      box.}
    \label{fig:basic2d}
\end{figure}

Figure \ref{fig:basic2d} summarizes the physical and chemical systems
associated with our control run that includes the reactive ozone
chemistry, described by reactions R1--R8. Below an altitude of ten
kilometres there are a number of differences between the day and night
sides of the tidally locked planet, as expected. Compared to the night
side, the day side wind speeds are $\simeq$20~m/s slower, specific
humidity is much higher and increases towards the surface, UV is
non-zero and falls off steadily as a function of depth associated with
cloud cover and ozone chemistry. The small, non-zero amount of
humidity on the nightside reflects transport from the dayside near the
terminator. Temperature is approximately constant from an altitude of
2~km towards the surface on the day side, while there is a strong
negative gradient on the night side over the same altitude
range. Above 10~km, there is little difference between the
distribution and values of air temperature and wind speed, suggesting
an efficient transport of heat between the hemispheres
\cite[]{Yang2014a, Lewis2018}. There is a strong peak in wind speed
between 20 and 30~km, reflecting a jet stream structure. This is in
broad agreement with \cite{Boutle2017}.

Figure \ref{fig:basic2d} also shows that ozone is vertically
distributed, peaking between 20 and 30~km; this is similar to the
structure we find in Earth's atmosphere. This peak in ozone is
slightly broader on the nightside, reflecting the presence of cold
traps that result in a local build-up of ozone. R1 and R4 describe the
production and loss of ozone. On the dayside there is a large
production rate due to the photolysis of O$_2$ throughout the
atmosphere, falling off towards the surface due to clouds; this
reflects the penetration of UV through the atmosphere. The production
rate is much smaller for the nightside, with non-zero values due to
transport from the dayside. The dayside loss rate (R4) is highest
above 10~km, some 90\% smaller than the production rate. The
corresponding nightside loss rate is much smaller, mainly due to the
absence of O($^1D$) produced by the photolysis of O$_3$. Below 10~km,
the dominant loss rates are due to the HO$_x$ catalytic cycle, with a
vertical distribution following the abundance of atmospheric humidity,
as expected. On the nightside, HO$_x$ radicals are exclusively
transported from the dayside, where the associated {\em in~situ}
initiation step R5 and the termination step R8 are extremely small
(Figure \ref{fig:basic2d}).

\begin{figure}
    \centering \includegraphics[width=\textwidth]{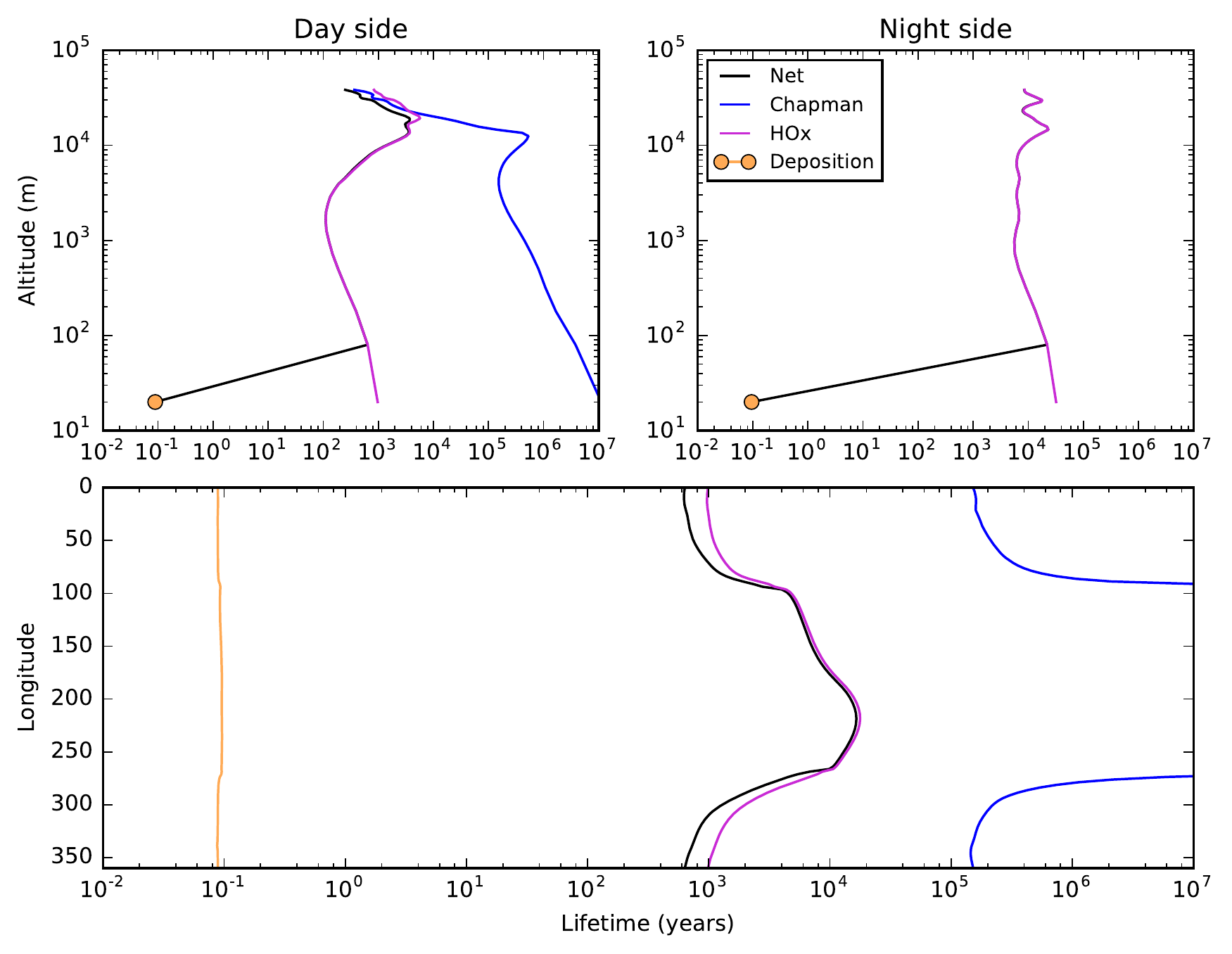}
    \caption{The total and contributing atmospheric lifetimes on a tidally
      locked M dwarf planet, expressed
      in years, of atmospheric ozone against chemistry and deposition on
      the (left panel) dayside and (right panel) nightside hemispheres.
      The bottom panel describes atmospheric lifetime as a function of
      longitude (mean taken over latitude and altitude) encompassing both hemispheres.
      The contributing lifetimes denoted
      as Chapman and HOx denote the lifetime of ozone against R4 and against
      the HOx catalytic cycle, respectively. Values correspond to a 120-day mean
      immediately succeeding a 20-year spin-up period.
    }
    \label{fig:lifetimepart}
\end{figure}

Figure \ref{fig:lifetimepart} shows the lifetime of atmospheric ozone
against the Chapman loss term, the HO$_x$ catalytic cycle, and from
deposition. We also show the mean net ozone lifetime corresponding to
the action of these three loss processes. This lifetime is of the
order of hundreds of years, increasing to 10,000 years when ozone is
trapped in the nightside cold traps, described below, with a sharp
increase in ozone lifetime at the terminators that reflects a rapid
decrease in UV radiation necessary to generate atomic oxygen (R1). We
find that the loss due to the HO$_x$ catalytic cycle largely
determines the net atmospheric lifetime of ozone on the day and
nightsides of the planet. The only exception is near the surface where
the deposition loss process dominates. The atmospheric lifetime of
ozone is an order of magnitude longer on the nightside of the planet,
as expected, due to the absence of photons.
Atmospheric transport and diffusion processes are orders of magnitude
faster than loss processes due to chemistry. The corresponding
residence times (non-chemistry lifetimes) typically range 0.1--10
hours in a grid box, and occasionally 100 hours in the laminae layers
of low-wind speed that lie between the jet streams, as discussed
below. 

\subsection{3-D Ozone Structure}

Here, we focus on the 3-D structure of atmospheric ozone and its
precursors, and refer the reader to Appendix \ref{app:A} for the
associated meteorological fields.

\begin{figure}
    \centering \includegraphics[width=\textwidth]{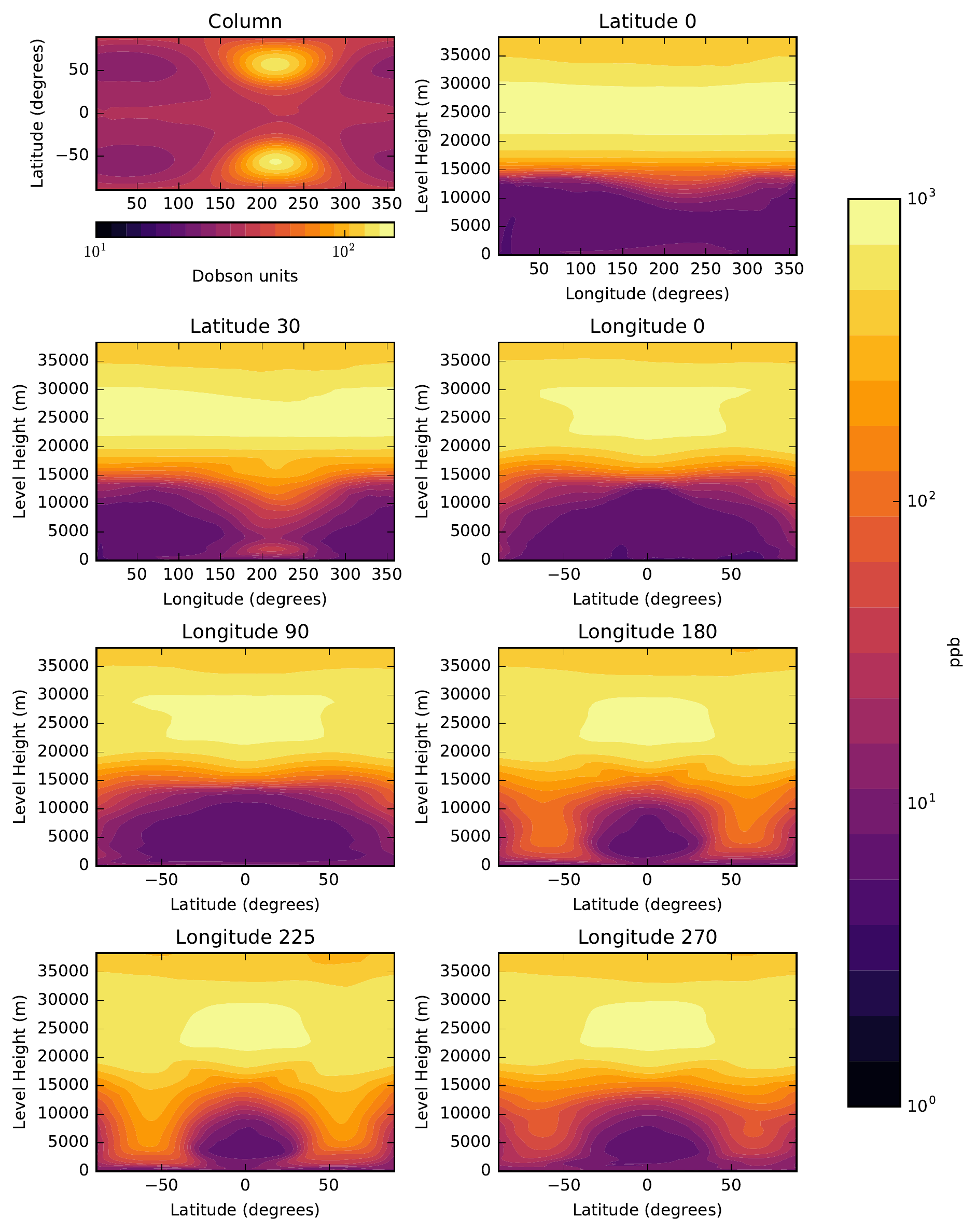}
    \caption{Atmospheric ozone distributions on a tidally-locked M dwarf
      planet. Values correspond to a 120-day mean immediately succeeding
      a 20-year spin-up period. The top left panel shows $\oz$ columns in
      Dobson units (DU), where one DU is equivalent to a layer of ozone
      that is 0.01~mm thick at standard temperature and pressure. The
      remaining panels show meridional or zonal slices that correspond
      to latitudes and longitudes, respectively, across the planet.}
    \label{fig:ozone}
\end{figure}

Figure \ref{fig:ozone} shows the 3-D distribution of atmospheric
ozone. We also show integrated column amounts of ozone, expressed in
Dobson units (DU) that are commonly used to report ozone in Earth's
atmosphere. One DU is defined as the thickness (in units of
10~microns) of the layer of a pure gas that would be formed by the
integrated column amount at standard temperature and pressure,
equivalent to 2.69$\times$10$^{20}$~molec/cm$^2$.  Ozone columns are
largest on the nightside, localized within two Rossby gyres either
side of an approximately uniform equatorial band that corresponds to
the rapid transport of air in the equatorial zonal jet
\cite[]{Showman2011}. The cyclonic gyres effectively trap air that
subsequently experience extended periods of radiative cooling,
resulting in localized columns of cold air. Within these cold traps
the atmosphere partially collapses (Figure \ref{fig:ozone}, longitude
225$^\circ$) so that large values of ozone, typical of the higher
atmosphere, are brought down to lower altitudes, resulting in higher
ozone columns. On the dayside, ozone columns are largest at the
poles. The smallest ozone columns ($\simeq$30~DU) are found the
dayside at northern and southern midlatitudes. Column ozone values on
our M dwarf planet are typically an order of magnitude smaller than
those found on Earth ($\simeq$300~DU) with the exception of the
interior of the nightside Rossby gyres, where ozone is 140~DU.

\begin{figure}
    \centering \includegraphics[width=\textwidth]{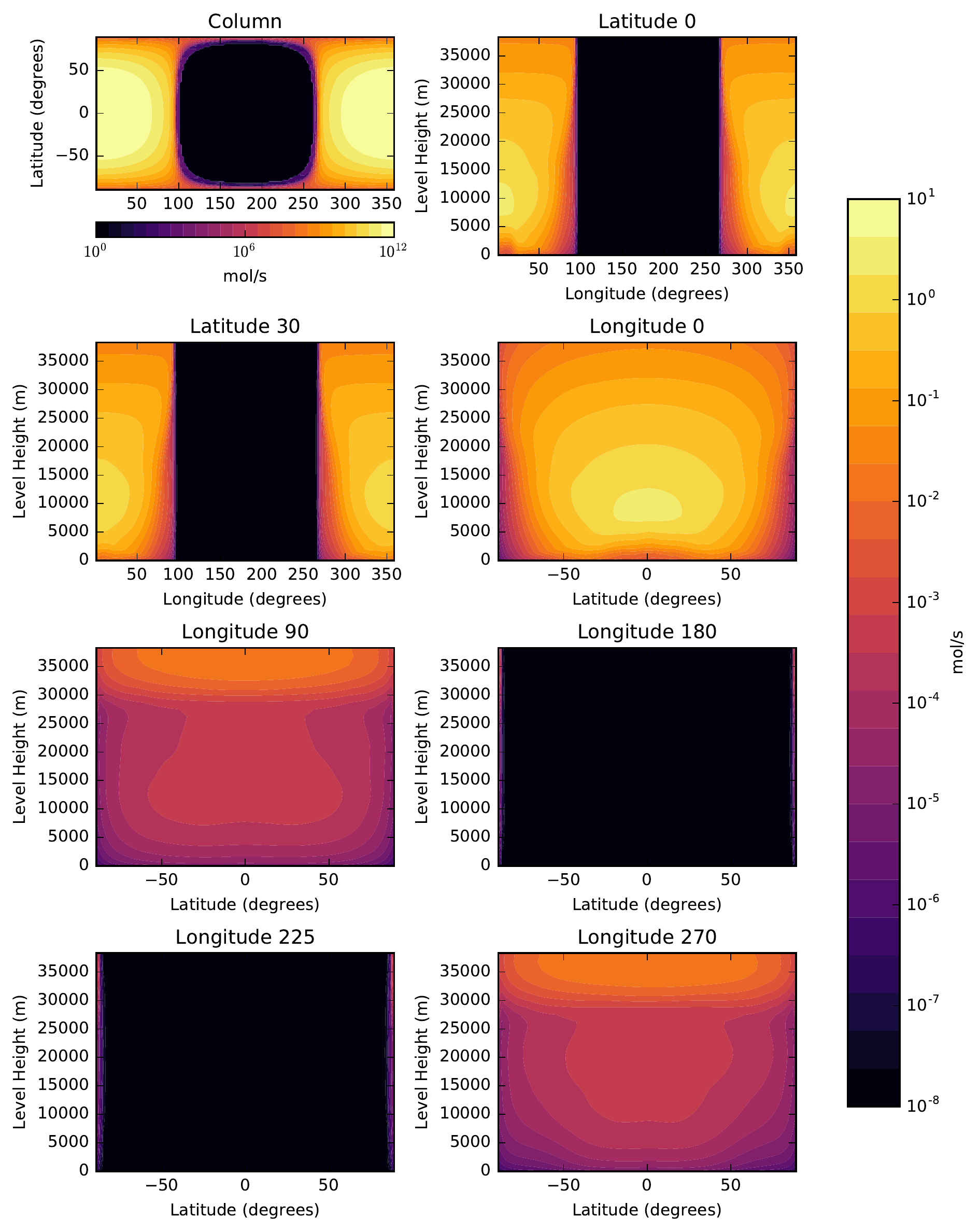}
    \caption{
      Reaction flux (mol/s) O$_2$ + h$\nu$ $\rightarrow$ O + O (reaction R1)
      corresponding to atmospheric ozone distributions on a tidally-locked M dwarf
      planet. Values correspond to a 120-day mean immediately succeeding a 20-year
      spin-up period. The format of the plot corresponds to Figure \ref{fig:ozone}.
    }
    \label{fig:R1}
\end{figure}

\begin{figure}
    \centering \includegraphics[width=\textwidth]{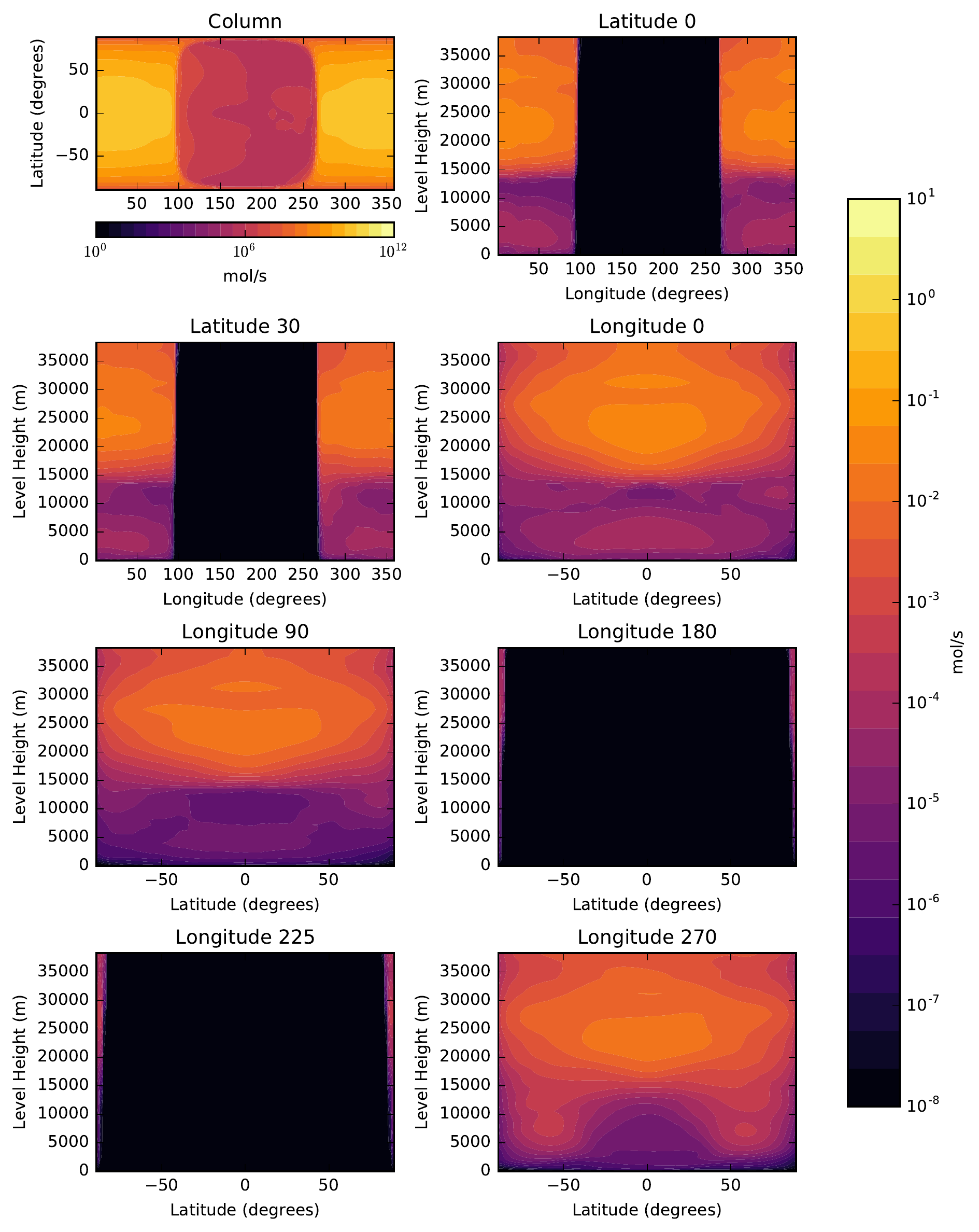}
    \caption{
      As Figure \ref{fig:R1} but for reaction flux (mol/s)
      O$_3$ + O $\rightarrow$ 2O$_2$ (reaction R4).
    }
    \label{fig:R4}
\end{figure}

\begin{figure}
    \centering \includegraphics[width=\textwidth]{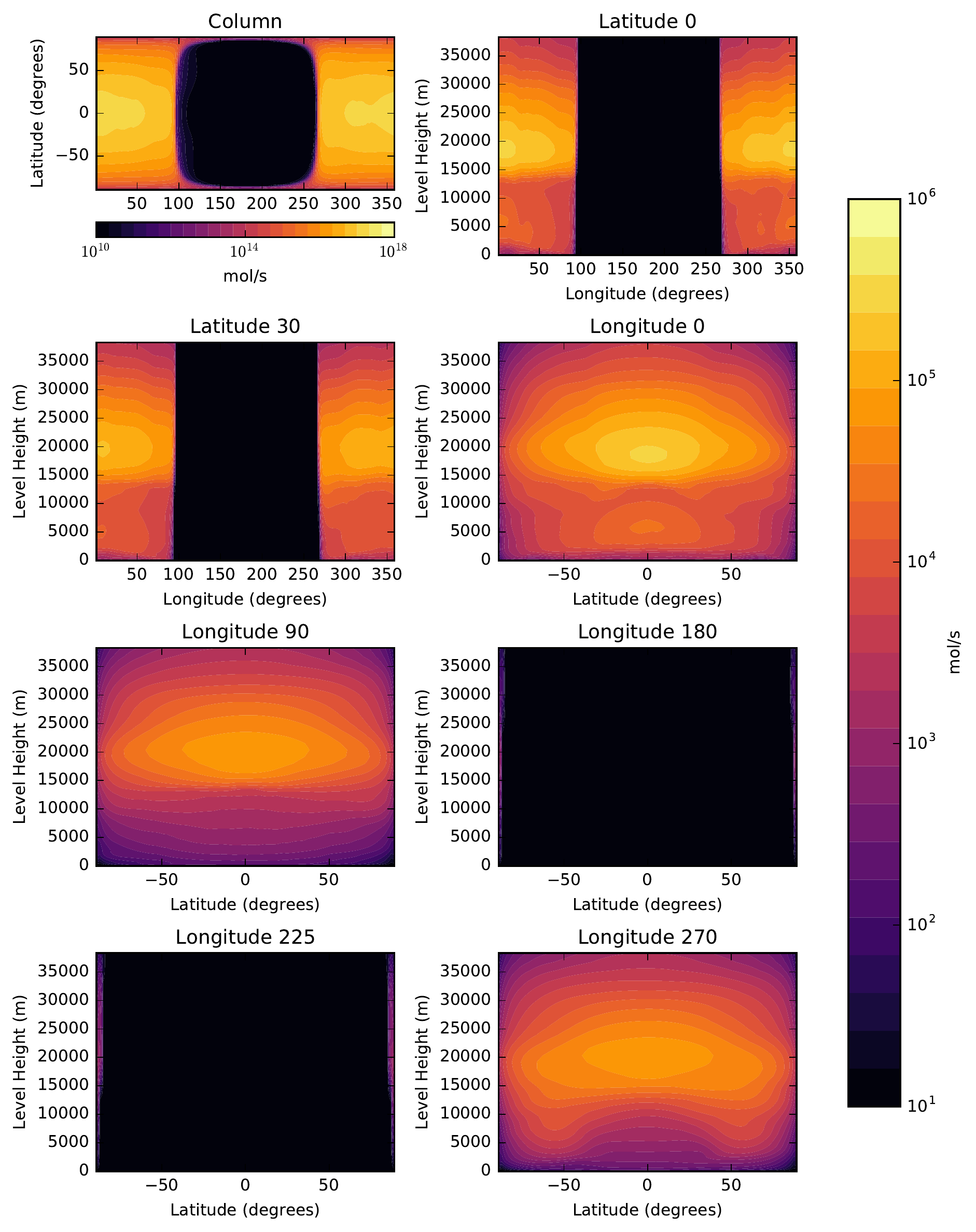}
    \caption{
      As Figure \ref{fig:R1} but for reaction flux (mol/s)      
      O + O$_2$ + H $\rightarrow$ O$_3$ + H (reaction R2).
    }
    \label{fig:R2}
\end{figure}

\begin{figure}
    \centering \includegraphics[width=\textwidth]{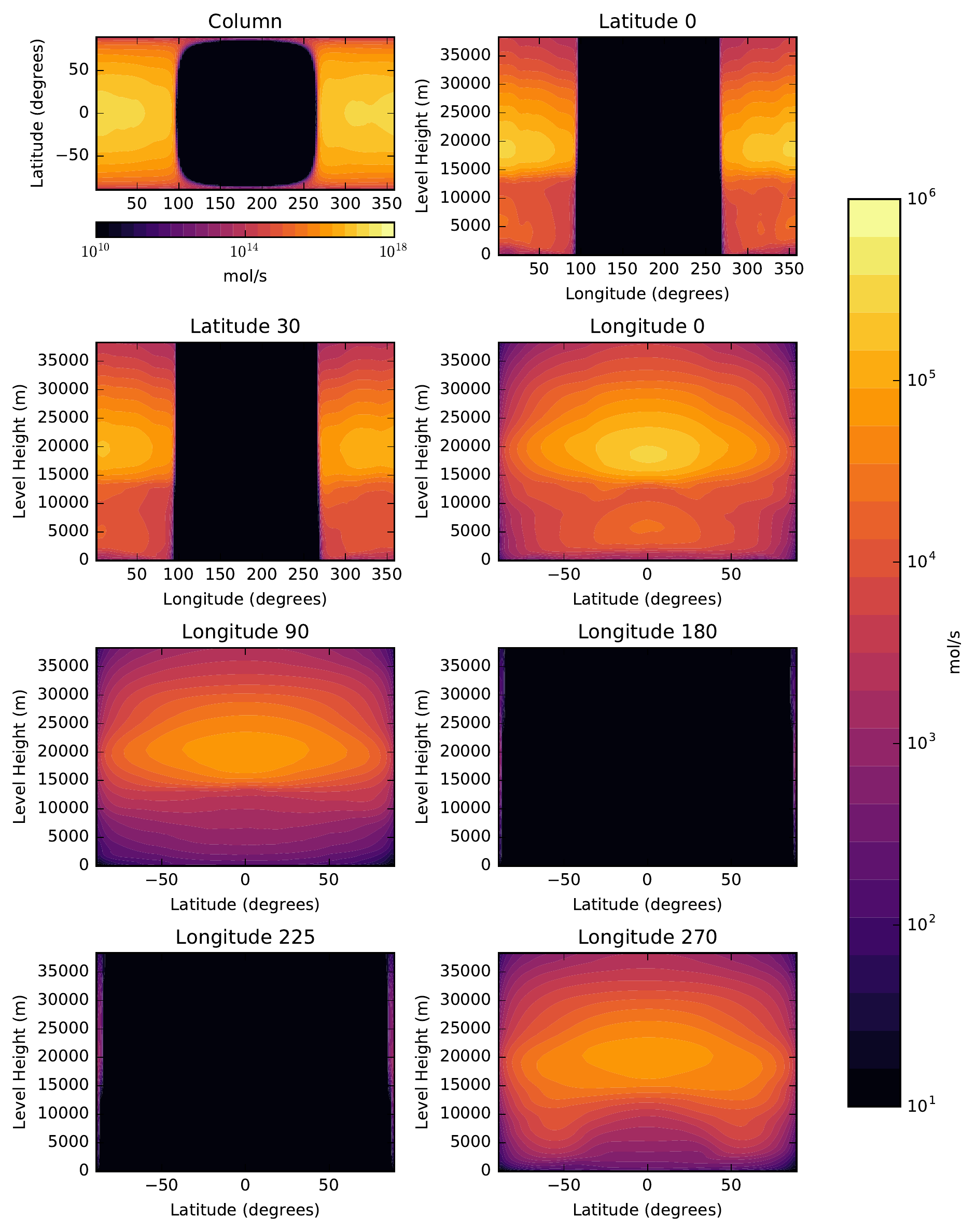}
    \caption{
      As Figure \ref{fig:R1} but for reaction flux (mol/s)            
      O$_3$ + h$\nu$ $\rightarrow$ O$_2$ + O (reaction R3).
    }
    \label{fig:R3}
\end{figure}

Figure \ref{fig:ozone} also shows atmospheric mole fractions of ozone
are at a minimum on the day side near the substellar point. This
corresponds to the location of the maximum ozone production and loss
rates (R1 and R4), Figures \ref{fig:R1} and \ref{fig:R4}), driven by
the availability of incoming UV radiation. The fast exchange between O
and ozone, R2 and R3 (Figures \ref{fig:R2} and \ref{fig:R3},
respectively), results in reaction fluxes that are up to five orders
of magnitudes larger than that the Chapman production (R1) and loss
(R4) fluxes of ozone and peaking on the dayside. In the absence of UV
radiation, nightside ozone mole fractions are determined mostly by
atmospheric transport but also by the comparatively slow HO$_x$
catalytic cycle loss process. We find that e-folding chemical
lifetimes of ozone (Figure \ref{fig:o3life}) within any grid box are
much longer than the residence time for an air mass to pass through
the same grid box (Figure \ref{fig:tlife}).  Initiation of the HO$_x$
catalytic cycle (R5) happens only on the day side, where warmer
atmospheric temperatures permit atmospheric humidity, but the
propagation reactions R6 and R7 (Figures \ref{fig:R6} and
\ref{fig:R7}) continue on the nightside. 

\begin{figure}
    \centering
    \includegraphics[width=\textwidth]{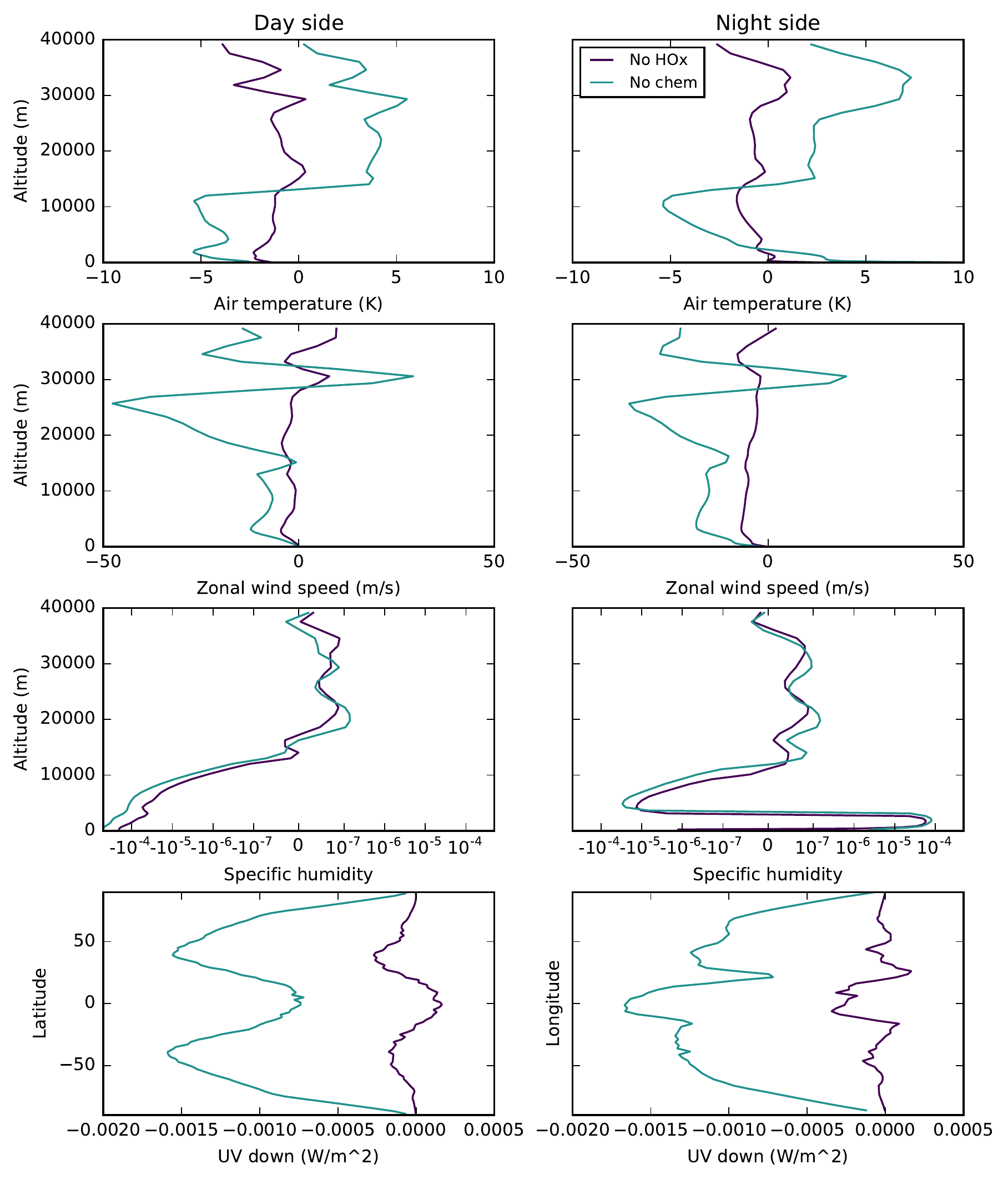}
    \caption{Sensitivity minus control differences of key meteorological
      parameters on a tidally-locked M dwarf planet that includes atmospheric
      ozone chemistry. The control calculation describes interactive ozone
      chemistry, described by the Chapman mechanism and the HO$_x$ catalytic
      cycle, that is consistent with the physical environment. The two sensitivity
      runs are 1) No-Chem, which uses a static Earth-like 3-D ozone distribution;
      and 2) No HO$_x$, which describes the interactive ozone chemistry but without
      the catalytic cycle. The top three panels describe (left) dayside and (right)
      nightside hemispheric mean values. For the specific humidity plot, the abscissa is
      symmetric about zero: linear between $-10^{-7}$ and $10^{-7}$) with
      the remainder described on a logarithmic scale. The bottom panels show zonal
      and meridional mean surface UV for the planet dayside.
    }
    \label{fig:key}
\end{figure}

To understand the impact of our assumed ozone chemistry mechanism on
the climate of Proxima Centauri b, we present model results from a
sensitivity run that uses an alternative fixed distribution of
Earth-like (larger) ozone that ranges from 2.4$\times$10$^{-8}$ to
1.6$\times$10$^{-5}$ MMR with the largest values in the stratosphere
\cite[]{Boutle2017}. We denote this run as ``No chem''. We also
consider a model run (``No HOx'') in which we remove the HO$_x$ ozone
catalytic cycle (reactions R5--R8).
Figure \ref{fig:key} shows the mean hemispheric dayside and nightside
values of the sensitivity minus control runs.  On the dayside,
``No-chem'' ozone results in 5~K warmer and 5~K cooler temperatures
above and below 15~km, respectively. Increased temperatures correspond
to the altitude of peak ozone due to ozone absorption and heating.  On
a hemispheric scale, atmospheric ozone reduces the amount of incoming
radiation penetrating through the atmosphere, with the largest
reductions at the sub-stellar point, thereby cooling altitudes
below. Larger Earth-like ozone values result in a near-surface cooling
of 2.6~K. We find a similar situation on the nightside, except that
near-surface temperatures are 4~K warmer, due to faster movement of
dayside polar air to midlatitudes on the nightside. These warmer
nightside temperatures are accompanied by higher levels of specific
humidity below 10~km.
Removing the HO$_x$ catalytic cycle from the interactive chemistry
scheme generally has a smaller impact on atmospheric physics than
adopting a fixed Earth-like ozone distribution. The catalytic cycle is
most important in the upper atmosphere of the sub-stellar point.
Figure \ref{fig:key} also show dayside surface UV zonal and meridional
mean distributions, which directly reflect the differences between the
control and sensitivity ozone distributions. 

\section{Discussion and Concluding Remarks}
\label{sect:discuss}

We use the Met Office Unified Model (UM) to explore the potential of a
tidally locked M dwarf planet, nominally Proxima Centauri b irradiated
by a quiescent version of its host star, to sustain an atmospheric
ozone layer. We have built on previous work \cite[]{Boutle2017} by
including the Chapman ozone photochemistry mechanism. We have also
included the hydrogen oxide (HO$_x$, describing the sum of OH and
HO$_2$) catalytic cycle to account for atmospheric water vapour that
plays a role in determining the distribution of atmospheric ozone.
We find that the host M dwarf star radiates with sufficient energy at
UV wavelengths to initiate and sustain an ozone layer on Proxima
Centauri b. The quasi stationary atmospheric distribution of
atmospheric ozone is determined by photolysis driven by incoming
stellar radiation and by atmospheric transport.  Ozone production
rates, determined by the photolysis of molecular oxygen, and loss
rates, determined by reaction between ozone and excited atomic oxygen,
are largest near the sub-stellar point, as expected. Reaction fluxes
involving photolysis get progressively smaller towards the
terminators, as expected. Concurrently, excited atomic oxygen is
quenched by molecular oxygen to rapidly form ozone, resulting in large
ozone concentrations on the nightside of the planet. We find the ozone
mole fractions are smallest (largest) in the lowest 15 km of the
atmosphere at the (anti) sub-stellar point. Above 15 km the ozone
distribution is dominated by an equatorial jet stream that rapidly
moves ozone around the planet. The nightside ozone distribution is
dominated by two Rossby gyres that leads to prolonged radiative
cooling of trapped air, resulting in a localized collapse of the
atmosphere that brings down higher ozone values to the lower
atmosphere. 

The calculations we report here are valid for a quiescent version of
the host star. Preliminary calculations (not shown) that crudely
describe a stellar electromagnetic flare as a large (150\%--600\%) and short
(1~hr--1~day) perturbation to the quiescent stellar flux received at
the top of our unmagnetized planetary atmosphere (Figure \ref{ch3:fig:spectrum})
show that the ozone layer is resilent. The larger flux of high-energy
photons on the dayside titrates O$_2$ (R1) and low-energy photons
titrate O$_3$ (R3) but can regenerate using products from elevated
photolysis fluxes and from reactions active on the nightside. This is
qualitatively consistent with 1-D calculations \cite[]{Segura2010}. More recent
1-D calculations \cite{Tilley2019} report that that if an electromagnetic
flare is accompanied by a proton ionization event, ozone can be permanently
removed from the atmosphere. The extent of this ozone removal is driven
by the magnitude, frequency, and duration of the stellar activity.

We find the planetary surface is potentially habitable in our
calculations, in agreement with \cite{Boutle2017}, with a significant
surface area with temperatures above 273~K. We did not consider
ice-albedo feedback, following \cite{Shields2013, Boutle2017,
  Lewis2018} who showed this effect is small for M dwarf planets.  For
our planetary calculations we adopted Earth-like CO$_2$
concentrations, but we acknowlege they are likely too low if there
exists a carbonate-silicate negative feedback cycle
\cite[]{Walker1981}.  This feedback mechanism has the potential to
increase the geographical region with surface temperatures above
273~K, depending on a number of factors, e.g. surface geology and
tectonic activity.

  We show that the thin
ozone layer (tens of Dobson units) on the dayside, supported by
incoming UV radiation, is sufficient to reduce the incoming UV by 60\%
from 40 to 0~km, as inferred by reduced photolytic production rates of
excited atomic oxygen as a function of atmospheric depth. Previous
studies that used a 1-D model found similar results \cite[e.g.,
][]{OMalley-James2017}.  Near-surface warming results in water cloud
formation that further reduces the UV penetration in the
atmosphere. Surface ozone levels are typically lower than 9~ppb, far
below values that are associated with values that trigger human
respiratory illnesses and crop damage.

A previous study \cite[]{Chen2018} that comes the closest to our
experiments is sufficiently different to permit only a qualitative
comparison. That study uses the CAM-Chem 3-D model for Earth with an
atmospheric chemistry scheme described by 97 species and 196
reactions, in comparison with our model with a reduced description of
ozone chemistry. The CAM-Chem model describes atmospheric dynamics
using 26 levels (surface--50~km) while our model uses 60 levels
(surface--80~km), allowing us to explore in more detail the general
circulation. They used stellar insolation of 1360~W~m$^{-2}$ while we
used 881.7 W~m$^{2}$, and our stellar spectrum contained less UV
radiation that even their quiescent M-dwarf. Finally,
\cite[]{Chen2018} used Earth continents as part of their surface
boundary conditions including a prescribe land/ice scheme, while we
use a shallow water world without continents or prescribed ice.
\citet{Lewis2018} showed that continents can significantly  affect the
climate of tidally-locked planets via convection and evaporation of
water, but acknowledge that \cite{Chen2018} placed their substellar
point in the middle of the Pacific Ocean where there are no large land
masses.  Despite these many differences between the two experiments,
our findings are broadly similar.  In particular, the contrast between
ozone on day and night sides is mainly determined by variations in UV
radiation, and corresponding changes in ozone lifetime, and the tidal
locking of the planet results increases the inhomogeneity of
ozone. Both studies also find that transported HO$_x$ species are an
important sink of night side ozone and the developing of a night side
temperature inversion.  However, there are some key differences in
ozone (compare their Figure 3 and our Figure \ref{fig:basic2d}) that
shows our ozone layer is much thinner than theirs, with our layer
thinning rapidly above 30~km while theirs extends above 48~km. It is
also unclear if their experiment results in the cold trap ozone
accumulation, which is a prominent night side feature. These
differences are likely due to the stellar spectral energy
distributions used and the role of continental orography affecting the
atmospheric dynamics.
 
A growing body of work has promoted the use of ozone as a potential
biosignature (\cite{Schwieterman2018} and references therein). In our
work we have shown that atmospheric ozone concentrations on a tidally
locked planet exhibit significant spatial variations across the
dayside hemisphere and between the dayside and nightside
hemispheres. This illustrates the need to model planetary atmospheres
in 3-D to more accurately reproduce atmospheric limb or disk-averaged
spectral measurements. The rapid ozone spatial variations we show at
the terminators, in particular, are relevant to transmission
spectroscopy along the atmospheric limb. A similar-sized planet in a
different orbit around a star, e.g. 3:2 resonance orbit, may very well
also exhibit strong temporal variations in ozone, reinforcing the need
to a) use 3-D models to describe ozone and other reactive chemistry
and b) observe the same objects repeatedly. We anticipate a more
realistic 3-D distribution of atmospheric constituents will help
reduce the probability of false positives and false negatives.

Our study has brought together researchers from Earth sciences and
astrophysics to improve understanding of exoplanetary atmospheres. We
have used a world-leading Earth system model that has been generalized
for exoplanets, providing us with a flexible numerical and scientific
framework on which to develop the physical, chemical, and potentially
biological processes necessary to understand the habitability of
individual exoplanets within the context of observational
constraints. Our work here has focused on the Chapman ozone mechanism
and a catalytic cycle associated with water vapour. Competing
catalytic cycles operate in Earth's atmosphere, some of which are
relevant to exoplanets. The nitrogen oxide catalytic cycle, initiated
through emission of NO$_x$ (biogenically or through geological
processes) or the thermal decomposition of N$_2$, can also lead to
efficient destruction of ozone. Lightning is one such process that could
achieve the necessary temperatures for thermal decomposition. More
generally, processes associated with a charged atmosphere can
significantly impact the physical and chemical state of the neutral
atmosphere. 

More broadly, we have shown that including a realistic description of
radiatively active atmospheric gases that is consistent with the
physical system significantly alters the physical state. Changing
atmospheric heating rates via absorption of incoming and outgoing
radiation alters a planet's ability to maintain liquid water at the
surface and therefore being considered habitable from an anthropic
perspective. Similarly, atmospheric particles lofted by aeolian
processes or from the condensation of non-volatile gases can scatter
and absorb incoming and outgoing radiation. To understand the
habitability of exoplanets, we have to understand their atmospheric
composition that can only be achieved through self-consistent global 3-D
models of atmospheric physics and chemistry.

\section*{Acknowledgements}

J.S.Y. was supported by the U.K. Natural Environment Research Council
(Grant NE/L002558/1) through the University of Edinburgh's E3 Doctoral
Training Partnership. P.I.P. gratefully acknowledges his Royal Society
Wolfson Research Merit  Award. We acknowledge Adam Showman for various
discussions and Paul Earnshaw who originally developed the slab ocean
model used in this simulation.  Materials produced using Met Office
Software. JM and IAB acknowledge the support of a Met Office Academic
Partnership secondment. We acknowledge use of the MONSooN2 system, a
collaborative facility supplied under the Joint Weather and Climate
Research Programme, a strategic partnership between the Met Office and
the Natural Environment Research Council. NJM is partly supported by a
Science and Technology Facilities Council Consolidated Grant
(ST/R000395/1), and a Leverhulme Trust Research Project Grant. 





\bibliographystyle{mnras}
\bibliography{newbib-rev}


\pagebreak

\appendix{}
\section{Physical Description of Numerical Experiments}
\label{app:A}

Here, we focus on the meteorological variables that complement the ozone
chemistry fields shown in the main paper. 

\begin{figure}
    \centering \includegraphics[width=\textwidth]{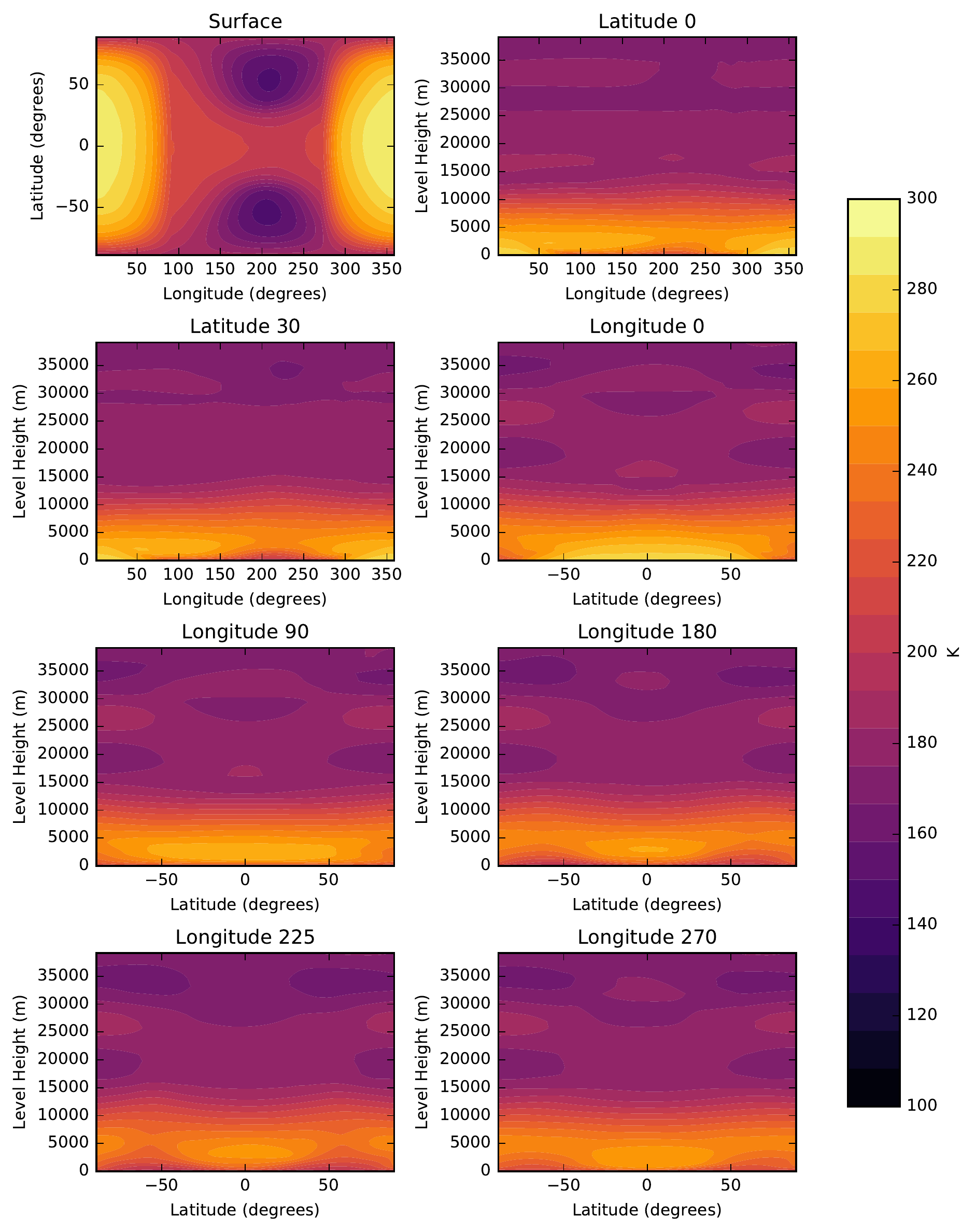}
    \caption{Air temperature in the simulated
      atmosphere. Panels show means over 120 days of model time. The
      top left panel shows surface temperature. Remaining panels show
      slices through latitudes $0^\circ$, $30^\circ$ and longitudes
      $0^\circ$, $90^\circ$, $180^\circ$, $225^\circ$ and
      $270^\circ$.}
    \label{fig:temp}
\end{figure}

\begin{figure}
    \centering \includegraphics[width=\textwidth]{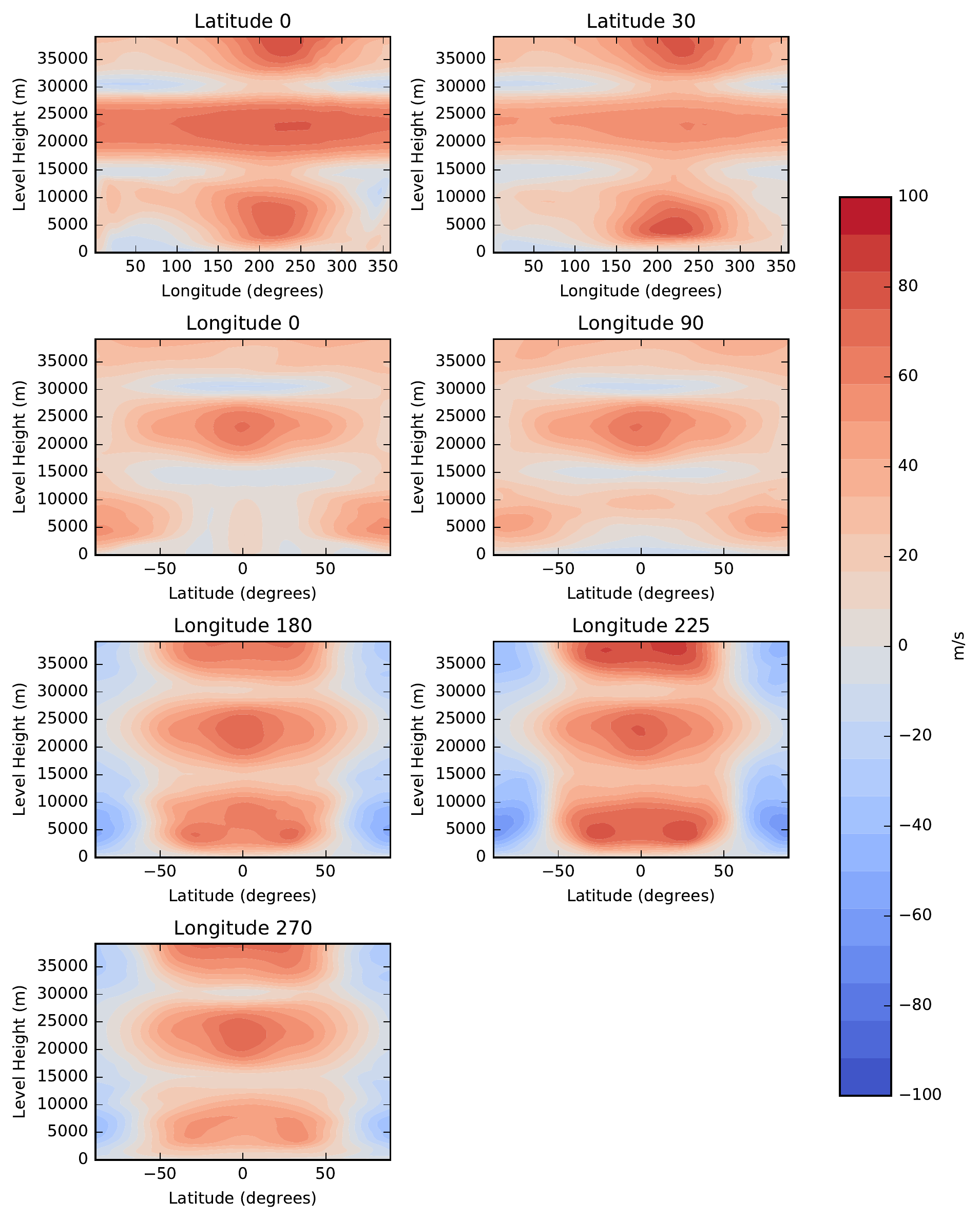}
    \caption{Zonal winds in the simulated
      atmosphere. Panels show means over 120 days of model
      time. Panels show slices through latitudes $0^\circ$, $30^\circ$
      and longitudes $0^\circ$, $90^\circ$, $180^\circ$, $225^\circ$
      and $270^\circ$.}
    \label{fig:uwind}
\end{figure}

\begin{figure}
    \centering \includegraphics[width=\textwidth]{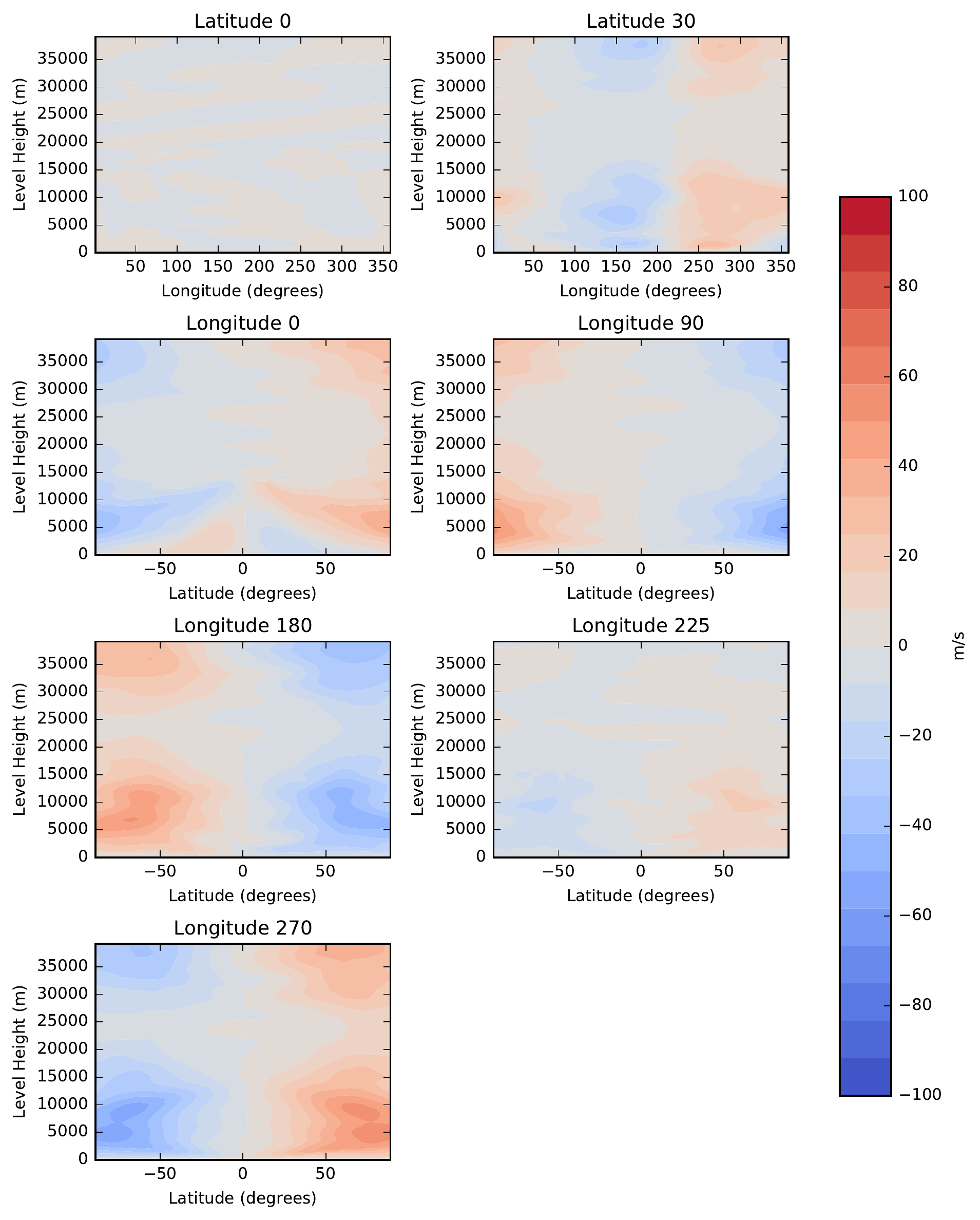}
    \caption{Meridional winds in the simulated
      atmosphere. Panels show means over 120 days of model
      time. Panels show slices through latitudes $0^\circ$, $30^\circ$
      and longitudes $0^\circ$, $90^\circ$, $180^\circ$, $225^\circ$
      and $270^\circ$.}
    \label{fig:vwind}
\end{figure}

Figure \ref{fig:temp} shows that surface temperatures on a significant
portion of the dayside hemisphere are warmer than the 273~K
triple point. Our calculations show that the planet can sustain liquid water at
the planetary surface \cite[]{Boutle2017}. On the nightside there are
symmetrically-placed cold traps at $\pm$50$^\circ$ latitude in the
northern and southern hemispheres, where temperatures are
$\simeq$150~K. Both cold traps are associated with cyclonic
movement. After the cold trap, air masses move towards polar latitudes
before rejoining the atmospheric jets at the anti-stellar point, where
there is a reversal in the meridional wind direction at approximately
at a longitude of 225$^\circ$ (Figures \ref{fig:uwind} and
\ref{fig:vwind}).  These cold traps result in longer, localized
residence times for air that impacts ozone chemistry (not
shown). These meteorological features are described further by
\cite{Showman2011}.

Maintenance of an atmospheric jet at 30~km (Figure \ref{fig:uwind})
provides a mechanism to warm the equatorial band on the nightside,
where atmospheric temperatures still remain below 273~K. We find a
second jet structure on the nightside below 10~km, with zonal winds
peaking close to the anti-stellar point (longitude
225$^\circ$). Between the jet streams on the day and night sides of
the planet we find layers where the wind speed is very low. Wind
speeds in the meridional direction (Figure \ref{fig:vwind}), driven by
the Coriolis force due to the planet orbiting its host star, are much
slower than zonal wind speeds that are driven by the thermal gradient
between the day and night sides. We also find a persistent thermal
inversion on the nightside with air between 2 and 5~km warmer than at
the surface. The maximum and minimum mean temperatures on the planet
are 287~K and 147~K, respectively. These mean extreme values are
$\simeq$3~K cooler than values consistent with Earth-like ozone
distributions \cite[]{Boutle2017}. 

\begin{figure}
    \centering \includegraphics[width=\textwidth]{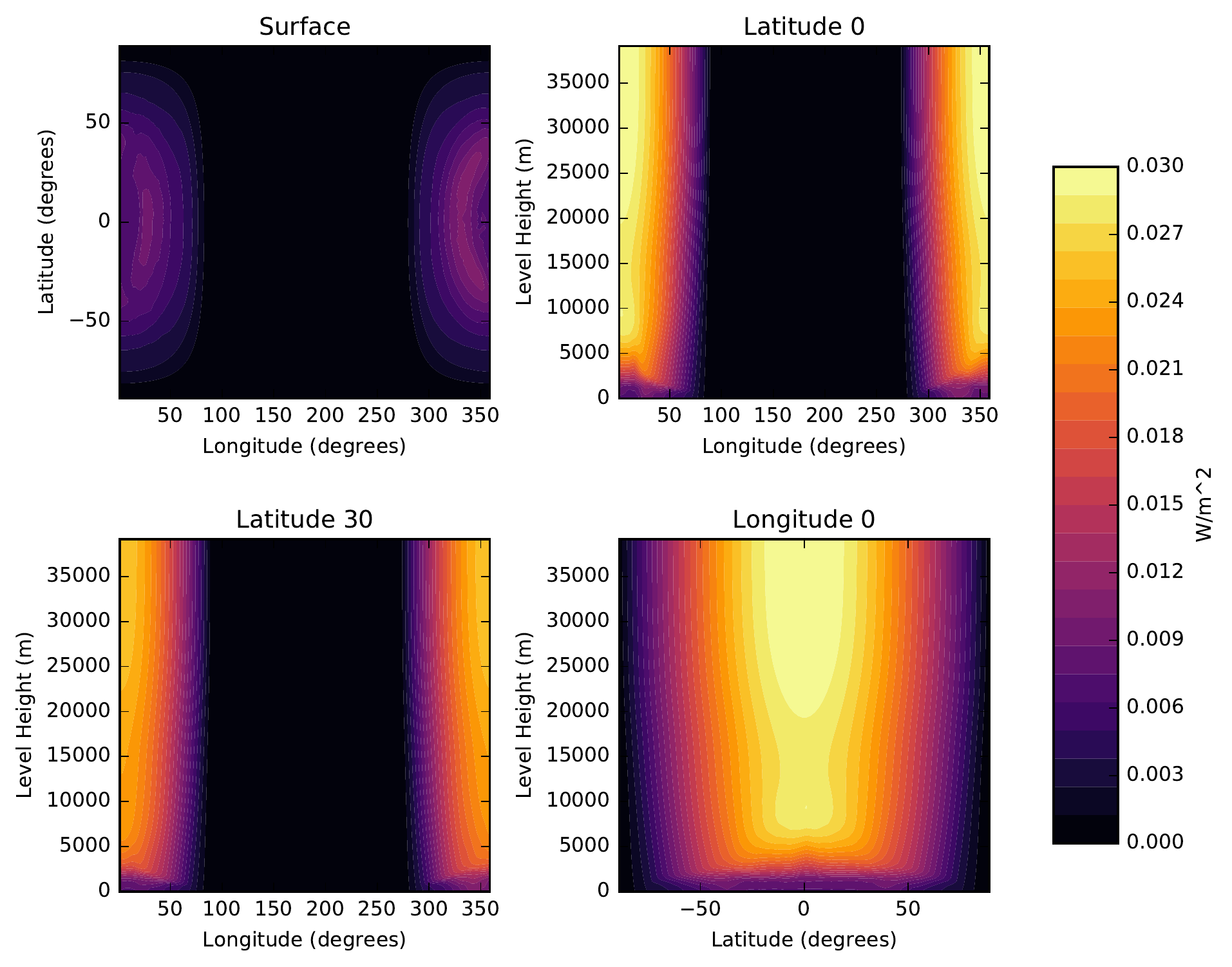}
    \caption{Downward UV in the simulated
      atmosphere. Panels show means over 120 days of model time. The
      top left panel shows surface UV. Remaining panels show slices
      through latitudes $0^\circ$, $30^\circ$ and longitudes
      $0^\circ$. We do not show the nightside plots here as there is
      obviously no UV flux at these locations.}
    \label{fig:uv}
\end{figure}

Figure \ref{fig:uv} shows the distribution of UV radiation as it
passes through the atmosphere. UV is non-zero on the dayside, as
expected, with smaller values towards the terminators. It is
relatively constant throughout the atmosphere, with a shallow gradient
below 30~km, due to ozone photolysis, followed by a sharp drop below
10~km due to cloud reflectivity that peaks at the substellar
point. This is similar to what happens on Earth. High-energy UVC is
rapidly absorbed at high altitudes, UVB is absorbed more slowly as it
passes through the stratospheric ozone layer, and UVA is attenuated at
lower altitudes but can also reach the surface. Our surface UV levels
are approximately 100 times lower than values we find on a similar
planet with an Earth-like top-of-the-atmospheric radiation flux using
the same planetary configuration but using a Sun-like spectral energy
distribution, a star-planet orbit of 1~AU, and a day length of 24~hrs.
This weaker level of incoming stellar UV radiation is still
sufficiently energetic to establish and maintain an atmosphere ozone
layer.



\section{Additional 3-D Chemistry Fields}
\label{app:B}

\begin{figure}
  \centering \includegraphics[width=\textwidth]{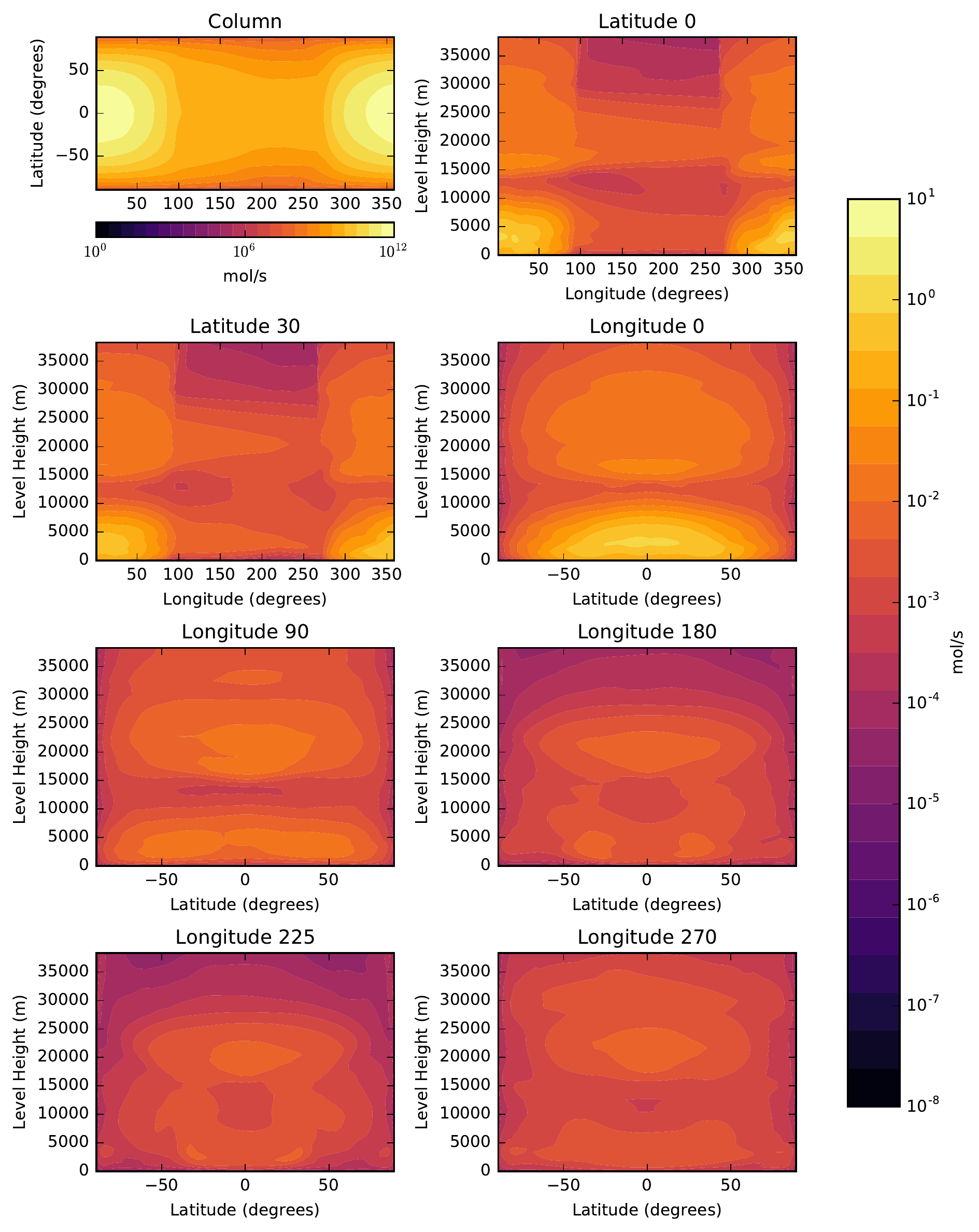}
  \caption{
    As Figure \ref{fig:R1} but for reaction flux (mol/s)              
    OH+O$_3$$\rightarrow$HO$_2$+O$_2$ (reaction R6).}
    \label{fig:R6}
\end{figure}

\begin{figure}
  \centering \includegraphics[width=\textwidth]{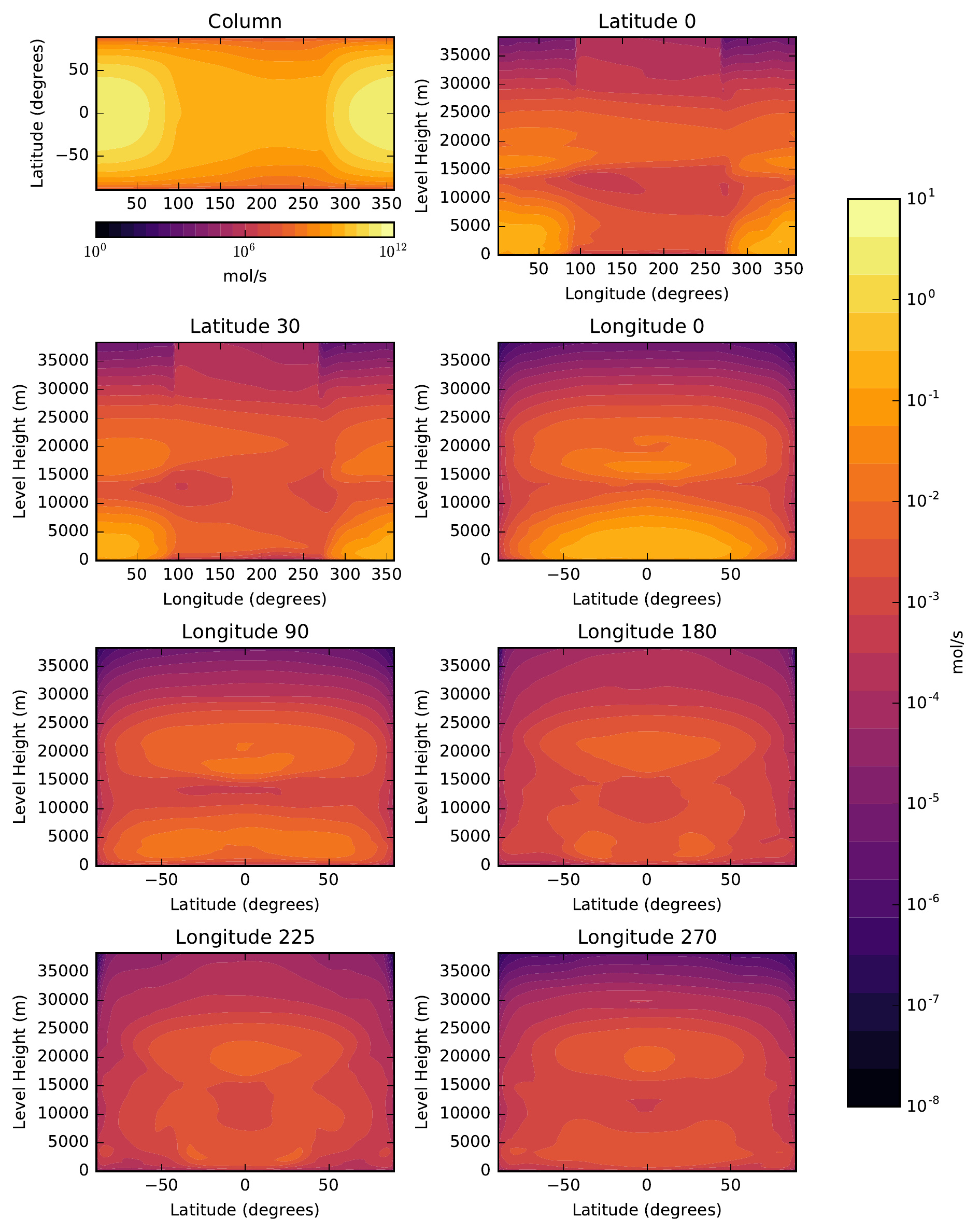}
  \caption{
    As Figure \ref{fig:R1} but for reaction flux (mol/s)              
    HO$_2$+O$_3$$\rightarrow$OH+2O$_2$ (reaction R7).}
    \label{fig:R7}
\end{figure}

\begin{figure}
    \centering \includegraphics[width=\textwidth]{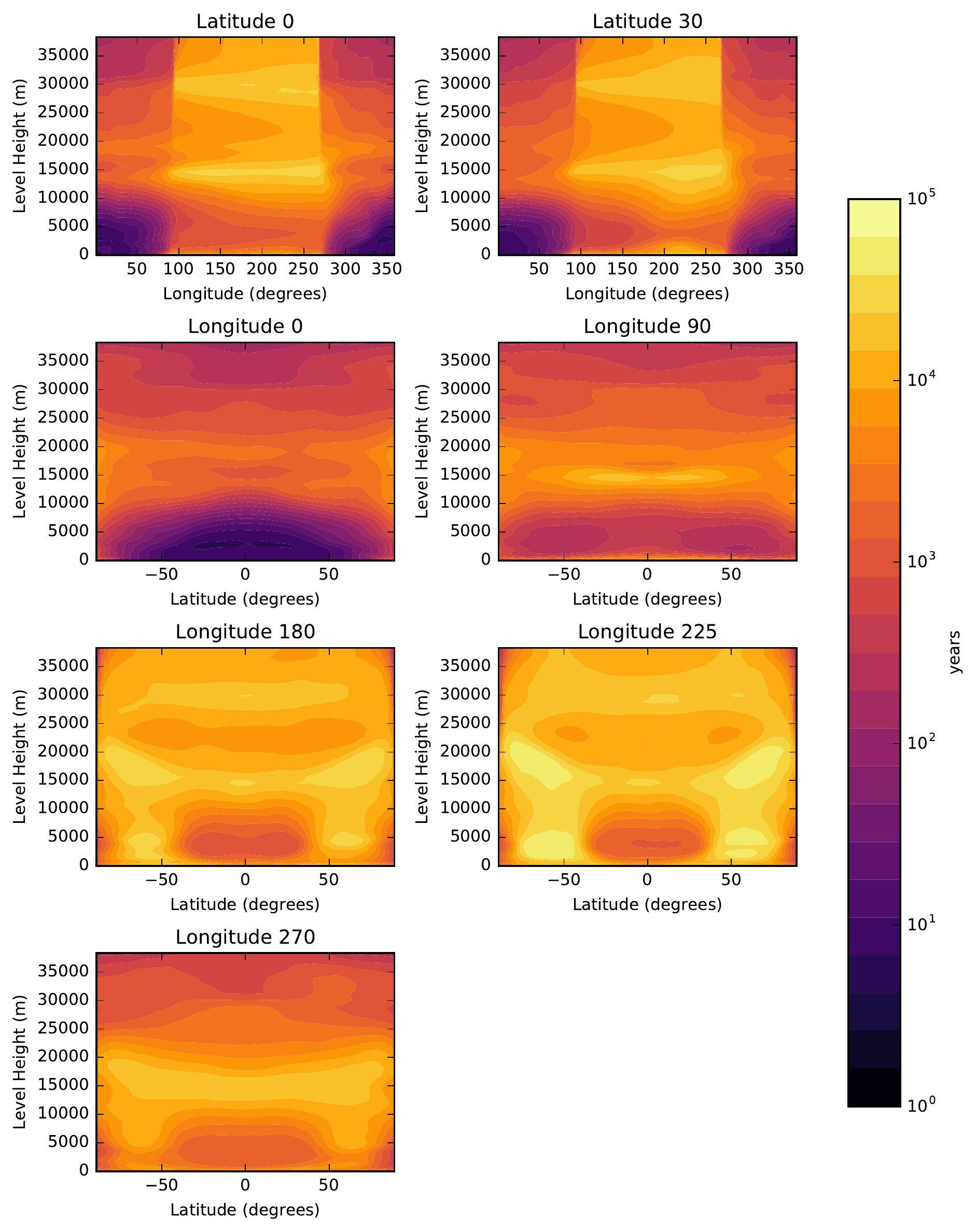}
    \caption[Net $\oz$ lifetime]{Net $\oz$ lifetime. Panels show means
      over 120 days of model time. Panels show slices through
      latitudes $0^\circ$, $30^\circ$ and longitudes $0^\circ$,
      $90^\circ$, $180^\circ$, $225^\circ$ and $270^\circ$.}
    \label{fig:o3life}
\end{figure}

\begin{figure}
    \centering \includegraphics[width=\textwidth]{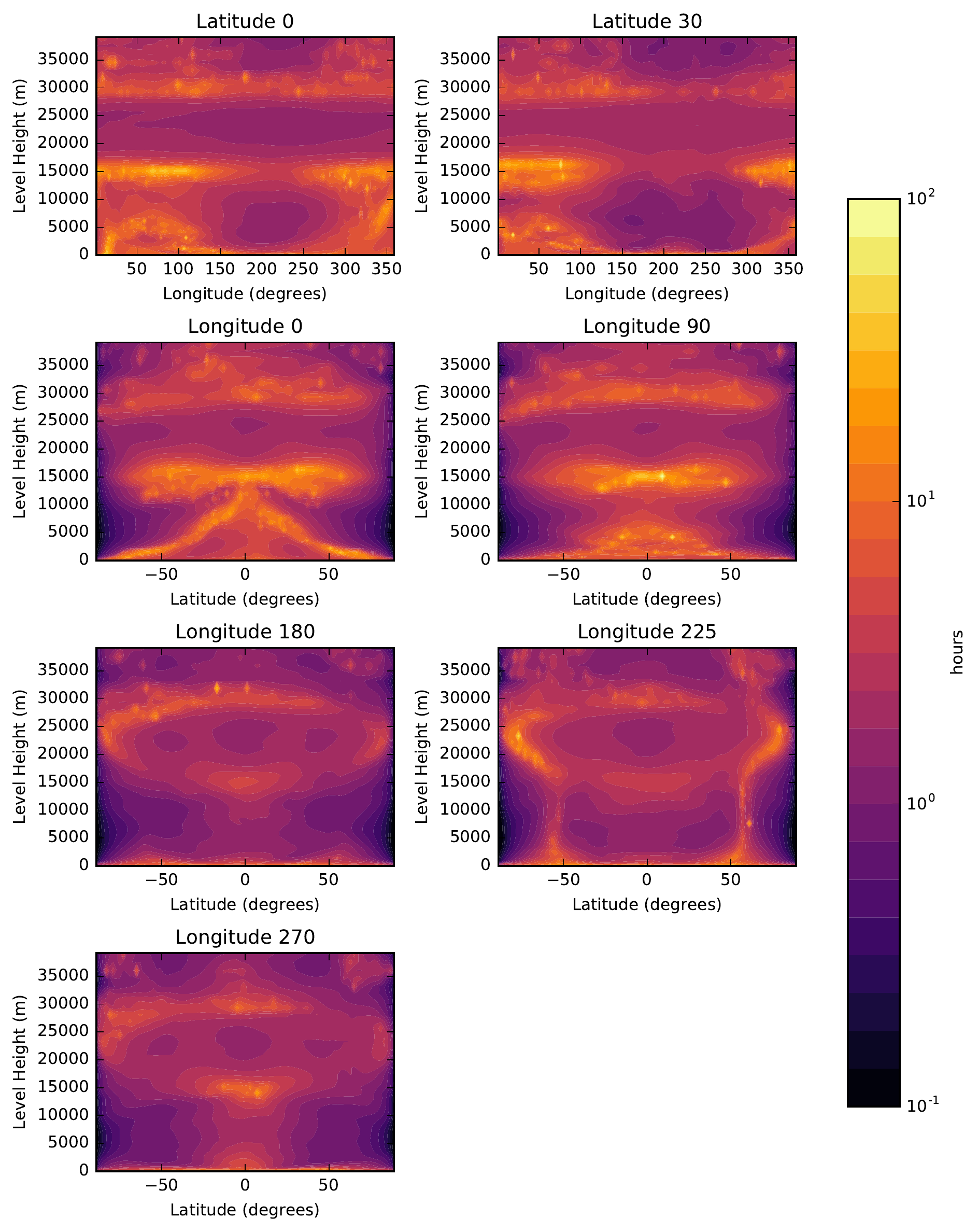}
    \caption[Transport lifetime]{Transport residence time for each grid
      box in the atmosphere. Panels show means over 120 days of model
      time. Panels show slices through latitudes $0^\circ$, $30^\circ$
      and longitudes $0^\circ$, $90^\circ$, $180^\circ$, $225^\circ$
      and $270^\circ$.}
    \label{fig:tlife}
\end{figure}

Figures \ref{fig:R6} and \ref{fig:R7} describe the key reaction fluxes associated
with the HO$_x$ catalytic cycle (reactions R5-R8 from the main text).
Figures \ref{fig:o3life} and \ref{fig:tlife} describe the atmospheric lifetime and
atmospheric residence time due
to chemical and transport processes, respectively.

%
%


\bsp	
\label{lastpage}
\end{document}